\documentclass{article}[12pt]
\usepackage[T1,T2A]{fontenc}
\usepackage[cp1251]{inputenc}
\usepackage{amsmath,amssymb}
\usepackage{graphicx}%
\usepackage{xcolor}
\usepackage{verbatim}
\usepackage{multirow}
\newtheorem{note}{Note}

\newcommand{\ThreeFig}[6]{%
\begin{flushleft}
\begin{tabular}{lcr}
\hspace{-1cm}
\parbox{5.5cm}{\includegraphics[width=5.5cm]{#1}}  & \hspace{-1.5cm} \parbox{5.5cm}{\includegraphics[width=5.5cm]{#2}}  & \hspace{-1.5cm}\parbox{5.5cm}{\includegraphics[width=5.5cm]{#3}}\\
\hspace{-1cm}
\parbox{5cm}{\vspace{7pt}\refstepcounter{figure} Fig. \thefigure.\quad #4\vfill} &\hskip -10pt \parbox{5cm}{\vspace{7pt}\refstepcounter{figure}Fig. \thefigure.\quad #5\vfill} & \parbox{5cm}{\vspace{7pt}
\refstepcounter{figure}Fig. \thefigure.\quad #6\vfill}\\
\end{tabular}
\end{flushleft}
\vspace{7pt}
}
\newcommand{\ThreeFigReg}[9]{%
\begin{flushleft}
\begin{tabular}{lcr}
\hspace{-1cm}
\parbox{5.8cm}{\includegraphics[width=5.45cm,height=#2cm]{#1}}  & \hspace{-0.8cm} \parbox{5.5cm}{\includegraphics[width=5.45cm,height=#4cm]{#3}}  & \hspace{-0.8cm}\parbox{5.5cm}{\includegraphics[width=5.45cm,height=#6cm]{#5}}\\
\hspace{-10pt}
\parbox{5cm}{\vspace{7pt}\refstepcounter{figure}Fig. \thefigure.\quad #7\vfill} &\hskip -10pt \parbox{5cm}{\vspace{7pt}\refstepcounter{figure}Fig. \thefigure.\quad #8\vfill} & \parbox{5cm}{\vspace{7pt}
\refstepcounter{figure}Fig. \thefigure.\quad #9\vfill}\\
\end{tabular}
\end{flushleft}
\vspace{7pt}
}
\newcommand{\TwoFigs}[4]{%
\begin{flushleft}
\begin{tabular}{cc}
\parbox{7cm}{\centerline{\includegraphics[width=5cm]{#1}}}  & \parbox{7cm}{\centerline{\includegraphics[width=5cm]{#2}}}  \\
\parbox{7cm}{\vspace{7pt}\refstepcounter{figure}Fig. \thefigure.\quad #3\vfill} & \parbox{7cm}{\vspace{7pt}\refstepcounter{figure}Fig. \thefigure.\quad #4\vfill}\\
\end{tabular}
\end{flushleft}
\vspace{7pt}
}
\newcommand{\TwoFigsReg}[6]{%
\begin{flushleft}
\begin{tabular}{cc}
\parbox{7cm}{\centerline{\includegraphics[width=6cm,height=#2cm]{#1}}}  & \parbox{7cm}{\centerline{\includegraphics[width=6cm,height=#4cm]{#3}}}  \\
\parbox{7cm}{\vspace{7pt}\refstepcounter{figure}Fig. \thefigure.\quad #5\vfill} & \parbox{7cm}{\vspace{7pt}\refstepcounter{figure}Fig. \thefigure.\quad #6\vfill}\\
\end{tabular}
\end{flushleft}
\vspace{7pt}
}
\begin{document}

\begin{center}
{\bf \Large Yu.G. Ignat'ev\footnote{Institute of Physics, Lobachevsky Institute of Mathematics and Mechanics, Kazan Federal University, Kremlyovskaya str., 35, Kazan, 420008, Russia; email: Ignatev-Yurii@mail.ru},  I.A. Kokh\footnote{Lobachevsky Institute of Mathematics and Mechanics, Kazan Federal University, Kremlyovskaya str., 35, Kazan, 420008, Russia; email: Irina\_Kokh@rambler.ru}} \\[12pt]
{\bf \Large Complete cosmological model based on a asymmetric scalar Higgs doublet} \\[12pt]
\end{center}

\abstract{A study of a complete cosmological model based on an asymmetric scalar doublet represented by the classical and phantom scalar Higgs fields is carried out. At the same time, the assumption about the nonnegativity of the expansion rate of the Universe, which in some cases contradicts the complete system of Einstein's equations, was removed. A closed system of dynamic equations describing the evolution of the cosmological model is formulated, and the dependence of the topology of the Einstein - Higgs hypersurface of the 5-dimensional phase space of the dynamical system that determines the global properties of the cosmological model on the fundamental constants of the model is investigated. A qualitative analysis of the dynamical system of the corresponding cosmological model is carried out, asymptotic phase trajectories are constructed, and the results of numerical modeling are presented, illustrating various types of behavior of the cosmological model.  
}

\section*{Introduction}
\par From a formal point of view, phantom fields appear to have been introduced into gravity as one of the possible scalar field models in 1983 by \cite{Ignat83_1}. In said work, as well as in later ones (see, for example, \cite {Ignat_Kuz84}, \cite{Ignat_Mif06}), phantom fields were classified as scalar fields with the attraction of like-charged particles and were identified by the factor $ \epsilon = -1 $ in  the energy - momentum tensor  of the scalar field.
 Note that phantom fields in the context to wormholes and the so-called black universes were considered in the works \cite{Bron1}, \cite{Bron2}. In this work, in accordance with the generally accepted terminology, we will call scalar fields with a negative kinetic term in the energy-momentum tensor phantom, regardless of the sign of the potential term. In this case, a negative potential term in the energy-momentum tensor corresponds to phantom scalar fields with the attraction of like-charged particles, and a positive potential term to phantom fields with repulsion. In the first case, the signs of the kinetic and potential terms coincide; in the second, they are opposite, which is equivalent to a change in the sign of the massive term in the Klein - Gordon equation. The corresponding solutions for a solitary scalar charge take not the form of the Yukawa potential, but the form of solutions of the equations of scalar Lifshitz perturbations for spherical symmetry ($\sin kr/r$) \cite{Ignat_12_3_Iz}.

\par However, the introduction of phantom fields into the structure of quantum field theory runs up against serious problems associated either with the probabilistic interpretation of quantum theory, or with the problem of the stability of the vacuum state, due to the unboundedness of negative energy \cite{Cline}. The negative kinetic term and violation of the isotropic energy condition imply that the energy is not bounded from below at the classical level, so negative norms appear at the quantum level. In turn, negative norms of quantum states generate negative probabilities that contradict the standard interpretation of quantum field theory \cite{Linde}, \cite{Saridakis}. The requirement that the theory be unitary leads to instability in describing the interaction of quantized phantom fields with other quantized fields \cite {Sbisa}.
In the paper \cite{Vernov}, however, it is noted that the terms leading to instability can be considered as corrections that are significant only at low energies below the physical cutoff. This approach allows us to consider the theory of phantom fields as some effective, physically acceptable theories, while it is assumed that an effective theory allows immersion in some fundamental theory, for example, field string theory, which is consistent with the well-known idea of S.M. Carroll, M. Hoffman and M. Trodden on effective field theory, in which the phantom model can be viewed as part or sector of the more fundamental \cite{Trodden} theory (see also \cite{Richarte}). In particular, in these works it was shown that there is a low-energy boundary, such that the phantom field will be stable during the lifetime of the Universe.

\par On the other hand, the analysis of observations of the cosmological acceleration, as well as the associated \ emph {barotropic coefficient} $w = p/ \varepsilon$,
carried out by various groups of researchers in recent years, shows that, apparently, it will be very difficult to do without phantom fields in cosmology. For example, SNIa data show a significant preference for "phantom" models and exclude the cosmological constant \cite {Tripathi}. Strict limits can be obtained in combination with other observational data, including measurements of the Hubble parameter $H(z)$ at different redshifts. When combining standard rulers and standard clocks, the best match is observed when $w_0 = -1.01 (+0.56 -0.31)$ \cite{Ma}. For the flat wCDM model, the constant parameter of the dark energy equation of state $w = -1.013 (+0.068 -0.073)$ \cite{Meyers} was measured, see also \cite{Terlevich, Chavez}.

\par Thus, in spite of all possible difficulties, cosmological models with phantom scalar fields have a right to exist. From a theoretical point of view, such models are very interesting and insufficiently studied. In this regard, the work \cite{Lazkoz} should be noted, in which, based on the study of the dynamical system of the quint, a counterexample of the typical behavior of the quint was presented, including attractors with $w\leqslant0 $. In the later, but independent works of one of the Authors with his students, the non-minimal theory of scalar interaction was consistently developed based on the concept of the fundamental scalar charge, both for classical and phantom scalar fields \cite{Ignat_12_1_Iz}, \cite{Ignat_12_2_Iz}, \cite{Ignat_12_3_Iz}. In particular, these works revealed some features of phantom fields, for example, features of interparticle interaction. Later, these studies were deepened to extend the theory of scalar, including phantom, fields to the sector of negative particle masses, degenerate Fermi systems, conformally invariant interactions, etc. \cite{Ignat_14_1_stfi}, \cite {Ignat_Dima14_2_GC}, \cite{Ignat_Agaf_Dima14_3_GC}, \cite{Ignat_15_1_GC}, \cite{Ignat_Agaf15_2_GC}. The mathematical models of scalar fields constructed in this way were applied to the study of the cosmological evolution of systems of interacting particles and scalar fields, both classical and phantom types \cite{Ignat_Mih15_2_Iz}, \cite{Ignat_Agaf_Mih_Dima15_3_AST}, \cite{Ignat_Agaf_16_3}. These studies revealed unique features of the cosmological evolution of plasmas with interparticle phantom scalar interaction, such as the existence of giant bursts of cosmological acceleration, the presence of a constant acceleration plateau, and other anomalies that sharply distinguish the behavior of cosmological models with a phantom scalar field from models with a classical scalar field. In particular, in the works \cite{Ignat_Agaf_16_3} - \cite{Ignat_Sasha_G&G}, a classification of types of behavior of cosmological models with interparticle phantom scalar field was carried out and 4 fundamentally different models were identified, among which there are models with $w\leqslant0$ at later stages evolution of the universe. The same papers also pointed out the possibility of Bose - condensation of nonrelativistic scalar charged fermions under conditions of a strong increase in the potential of the scalar field and considering this condensate as a component of dark matter. It is important to note the fact that in the case of a phantom field with attraction in the course of cosmological evolution, values of cosmological acceleration are attainable greater than 1, which, according to the generally accepted classification, corresponds to the phantom state of matter.

\par These studies show the need for a deeper study of both classical and phantom scalar fields with self-action as a possible basis for the cosmological model of the early Universe. In the works \cite{Ignat_16_5_Iz} - \cite{Ignat_Agaf_2017_2_GC}, a preliminary qualitative analysis of the cosmological model based on a phantom scalar field with self-action was carried out. In this paper, we will develop and detail the results of studies of cosmological models based on classical and phantom scalar fields. Unlike the works \cite{Ignat_14_1_stfi} - \cite{Ignat_Agaf_Mih_Dima15_3_AST} we will not take into account the contribution of ordinary matter, that is, we will consider free classical and phantom fields without a source.

In a number of works by I. Ya. Aref'eva, S. Yu. Vernov, A. S. Koshelev, et al. \cite{Vernov1} - \cite{Vernov3}, two-field cosmological models based on a pair of scalar fields, classical and phantom, in which a negative kinetic term corresponds to the phantom field. In this case, under the assumption of the polynomial potential of the 6th order scalar fields, the classes of one-parameter and two-parameter exact solutions were found. In particular, in S. Yu. Vernov's work \cite{Vernov3}, these solutions are found by the \emph{superpotential} method. As noted in \cite{Vernov2}, \cite{fantom} phantom fields are involved in cosmology to provide the value of the \emph{barotropic} coefficient $w<-1$ ($w: p = w \varrho$) in order to prevent ``the big break''. Note that in these papers, two-field cosmological models are substantiated by string theory, in which the tachyon describing brane decay is considered as a phantom field.

On the other hand, in a number of later but independent works \cite{Yu_17_1} - \cite{Yu_19_3}, the methods of the qualitative theory of dynamical systems and numerical-analytical methods were used to study cosmological models based on scalar fields, both with a quadratic potential and with Higgs potential. In particular, in the papers \cite{Yu_19_1, Yu_19_2}, two-field cosmological models, called in these papers the \emph{asymmetric scalar doublet}, were investigated. In these works, scalar fields were not tied to specific field theory models, therefore the fundamental constants of the corresponding cosmological models were arbitrary. However, this group of works used the condition the nonnegativity of the Hubble constant
\begin{equation}\label{H>=0}
H(t)\geqslant 0.
\end{equation}

The results of the above works led the Authors to the assumption about the possibility of the occurrence of the so-called \emph{limit Euclidean cycles} \cite{Yu_19_2}. The essence of Euclidean cycles lies in the tendency of cosmological models with certain parameters of the field model to a state with zero effective energy. In this case, space-time becomes pseudo-Euclidean, although the scalar fields are nonzero and perform free nonlinear oscillations. Such vacuum oscillations were studied in detail in the work \cite{Yu_19_3}.

In a private conversation with one of the Author on the sidelines of the conference on gravity, cosmology and astrophysics of the BRICS countries (Kazan, September 2019), Sergei Vernov expressed the opinion that the system of dynamic equations considered in the works \cite{Yu_17_1} - \cite{Yu_19_3} describing cosmological evolution may not be equivalent to the original Einstein system of equations. More precisely, the restriction of the Hubble constant to non-negative values \eqref{H>=0} may contradict the original system of Einstein's equations. This useful discussion resulted in the article \cite{Yu_20_1}, in which it was shown that the condition \eqref{H>=0} for $ H\to 0 $ actually contradicts the original system of Einstein-Higgs equations in the case of the classical Higgs field. Based on the formulated complete system of Einstein - Higgs equations\footnote{By the complete cosmological model we mean a model with the condition of nonnegativity of the Hubble constant \eqref{H>=0} removed.} In \cite {Yu_20_1}, a qualitative and numerical-analytical study of the original the system of dynamical equations for the \emph{classical} Higgs field, and the possibility in a number of cases of changing the expansion phase to the compression phase in the corresponding cosmological model is shown. Thus, the removal of the nonnegativity condition for the Hubble constant \eqref{H>=0} opened the Pandora's box and created new problems for cosmological models based on scalar fields.

We note a similar independent work in 2018 by G. Leon, A. Paliathanasis and J. L. Morales  \cite{Leon18}, in which a detailed analysis of the dynamical system corresponding to the quintom cosmological model with exponential potential energy of the classical and phantom scalar fields was carried out using the methods of the qualitative theory of differential equations $ V(\phi, \psi) = V_0 \exp (- \sqrt{6} (m \varphi + n \psi)) $ for cosmological models with different three-dimensional curvatures. In recent years, such models of scalar fields have been actively studied due to, first, the analyticity of the potential of scalar fields, and secondly, the possibility in a number of cases of constructing exact solutions, as well as a simple model of interaction between the components of the quint..

In this article, we will investigate a complete cosmological model based on an asymmetric scalar doublet. In this case, we will focus on the properties of the model, including its physical characteristics and the possibility of changing the phases of cosmological expansion and contraction. The need for such a study is dictated, among other things, by the fact that the classes of exact solutions obtained in the works \cite{Vernov1} - \cite{Vernov2}, firstly, are one-parameter or two-parameter, while the general solution of dynamic equations for a two-field model, in general speaking, for given values of the fundamental constants of the field model, it should be 5-parametric. Thus, there is no certainty that the solutions obtained belong to the class of solutions physically significant for cosmology. In short, these solutions lie on a two-dimensional surface in the five-dimensional phase space of the dynamical system of the cosmological model. In addition, the fundamental constants in the potentials of the considered two-field models are not arbitrary, reflecting the specifics of the string field theory model tied to these cosmological models.

In the course of this research, we will, first, widely use the concept of \emph{Einstein-Higgs hypersurface} introduced in \cite{Yu_20_1}, the topology of which determines the global behavior of the dynamical system. Second, we will apply the methods of the qualitative theory of dynamical systems (see, for example, \cite{Bogoyav}) to determine the asymptotic behavior of the cosmological model for arbitrary values of the parameters of the field model. Finally, thirdly, we will apply methods of numerical - analytical modeling in the system of computer mathematics.\footnote {This work was supported by the Kazan Federal University at the expense of the state assignment in the field of scientific activity.}


\section{Basic relations of the mathematical model}
\subsection{Field equations}
{As a field model, consider the self-consistent system of Einstein equations and the scalar doublet $\{\Phi,\varphi\}$ with the Higgs potential, which corresponds to the classical action
\begin{equation}\label{act}
S=-\frac{1}{8\pi}\int\biggl(\frac{1}{2}R+\Lambda+L\biggr)dV,
\end{equation}
where $\Lambda$ -- cosmological constant, $L$ -- Lagrange function of scalar doublet $\{\Phi,\varphi\}$ of noninteracting \emph{classical field} $\Phi$ and \emph{phantom field} $\varphi$
\begin{eqnarray} \label{Ls}
L=\frac{1}{16\pi } (g^{ik} \Phi _{,i} \Phi _{,k} -2V(\Phi ))-
\frac{1}{16\pi } (g^{ik} \varphi _{,i} \varphi _{,k} +2v(\varphi )),
\end{eqnarray}
where
\begin{eqnarray}
\label{Higgs}
V(\Phi )=-\frac{\alpha }{4} \left(\Phi ^{2} -e\frac{m^{2} }{\alpha } \right)^{2} ;\,\,\,\,
v(\varphi )=-\frac{\beta }{4} \left(\varphi ^{2} - \varepsilon\frac{\mathfrak{m}^{2} }{\beta } \right)^{2}
\end{eqnarray}
-- potential energy of the corresponding scalar fields, $\alpha$ and $\beta$ -- their self-action constants, $m$ and $\mathfrak{m}$ -- their masses of quanta, $e=\pm 1,\ \varepsilon=\pm 1$ -- indicators.

Using the standard procedure, we obtain from the Lagrange function \eqref{Ls} the equations of scalar fields
\begin{equation}\label{Eq_s}
\square \Phi +V'(\Phi )=0; \quad
\square \varphi -v'(\varphi)=0.
\end{equation}
and the energy-momentum tensor %
\begin{eqnarray}\label{T_{iks}}
T_{ik} =\frac{1}{16\pi } (2\Phi _{,i} \Phi _{,k} -g_{ik} \Phi _{,j} \Phi ^{,j} +2V(\Phi )g_{ik} )
-\frac{1}{16\pi } (2\varphi _{,i} \varphi _{,k}-g_{ik} \varphi _{,j} \varphi ^{,j} -2v(\varphi )g_{ik} ).
\end{eqnarray}
Note that the constant term in the Higgs potentials \eqref{Higgs} in the energy-momentum tensor leads to renormalization of the seed cosmological constant $\Lambda$:
\begin{equation}\label{L->L}
\Lambda_0\to \Lambda=\Lambda_0-\frac{m^4}{4\alpha}-\frac{\mathfrak{m}^4}{4\beta}.
\end{equation}

The corresponding Einstein equations of the system under study have the form \footnote{In this article, everywhere $G = \hbar = c = 1 $, the signature of the metric is $(-, -, -, +)$, the Ricci tensor is obtained by the convolution of the first and third indices.}:
\begin{eqnarray}\label{Eq_Einst}
G^i_k\equiv R^i_k-\frac{1}{2}R\delta^i_k =8\pi T^i_k+\Lambda_0\delta^i_k.
\end{eqnarray}
Equations \eqref{Eq_s} and (\ref{Eq_Einst}) are the basic model of the cosmological scenario.
\subsection{Dynamic equations for the spatially flat Friedmann model}
In the case of the spatially flat Friedmann Universe
\begin{equation}\label{Freed}
ds^2=dt^2-a^2(t)(dx^2+dy^2+dz^2)
\end{equation}
complete system of dynamic equations for the scale factor $a(\eta)$ and scalar potentials $\{\Phi(\eta),\varphi(\eta)\}$ takes the form \footnote{In these equations, the cosmological constant is renormalized according to \eqref{L->L}.}:
\begin{eqnarray}\label{Eq_Phi0}
\ddot{\Phi}+3\frac{\dot{a}}{a}\dot{\Phi}+e m^2 \Phi-\alpha\Phi^3=0;\\
\label{Eq_phi0}
\ddot{\varphi}+3\frac{\dot{a}}{a}\dot{\varphi}-\varepsilon \mathfrak{m}^2 \varphi+\beta\varphi^3=0;\\
\label{EqEinst0_4}
3\frac{\dot{a}^2}{a^2}-\frac{\dot{\Phi}^2}{2}-\frac{e m^2\Phi^2}{2}+\frac{\alpha\Phi^4}{4}
+\frac{\dot{\varphi}^2}{2}-\frac{\varepsilon \mathfrak{m}^2\varphi^2}{2}+\frac{\beta\varphi^4}{4}- \Lambda=0;\\
\label{EqEinst0_1}
2\frac{\ddot{a}}{a}+\frac{\dot{a}^2}{a^2}+\frac{\dot{\Phi^2}}{2}-\frac{e m^2\Phi^2}{2}+\frac{\alpha\Phi^4}{4}
-\frac{\dot{\varphi}^2}{2}-\frac{\varepsilon \mathfrak{m}^2\varphi^2}{2}+\frac{\beta\varphi^4}{4}-\Lambda=0.
\end{eqnarray}
Not all equations from the system of four differential equations \eqref{Eq_Phi0} - \eqref{EqEinst0_1} are independent. Differentiating the equation \eqref{EqEinst0_4} with respect to the time variable, we obtain the equality:
\begin{eqnarray}
\frac{3\dot{a}}{a}\biggl(2\frac{\ddot{a}}{a}-2\frac{\dot{a}\ \!^2}{a^2}+\dot{\Phi}^2-\dot{\varphi}^2\biggr)
-\dot{\Phi}\biggl[\ddot{\Phi}+3\frac{\dot{a}}{a}\dot{\Phi}+e m^2 \Phi-\alpha\Phi^3\biggr]
+\dot{\varphi}\biggl[\ddot{\varphi}+3\frac{\dot{a}}{a}\dot{\varphi}-\varepsilon \mathfrak{m}^2 \varphi+\beta\varphi^3\biggr]=0.
\end{eqnarray}
Due to the field equations \eqref{Eq_Phi0} - \eqref{Eq_phi0} we get from here:
\begin{equation}\label{a'(11-44)}
\frac{\dot{a}}{a}\biggl(\frac{\ddot{a}}{a}-\frac{\dot{a}\ \!^2}{a^2}+\frac{1}{2}\dot{\Phi}^2-\frac{1}{2}\dot{\varphi}^2\biggr)=0.
\end{equation}
Substituting into \eqref{a'(11-44)} the expression for $(-3\dot{a}^2/a^2)$ from the Einstein equation \eqref{EqEinst0_4} for $\dot{a}\not\equiv0$, we get the Einstein equation \eqref{EqEinst0_1}. Thus, the Einstein equation for the spatial components $Х^\alpha_\alpha$ \eqref{EqEinst0_1}  \textbf{for $\dot{a}\not\equiv0$} is a differential - algebraic consequence of the field equation \eqref{Eq_Phi0} and Einstein's equations for the $^4_4$ component. Therefore, the system of equations \eqref{Eq_Phi0} -- \eqref{Eq_phi0} and the difference of equations \eqref{EqEinst0_1} -- \eqref{EqEinst0_4} can be considered as the basic dynamical system of a spatially flat cosmological model:
\begin{equation}\label{(11-44}
\frac{\ddot{a}}{a}-\frac{\dot{a}\ \!^2}{a^2}+\frac{1}{2}\dot{\Phi}^2-\frac{1}{2}\dot{\varphi}^2=0.
\end{equation}
Introducing  the \emph{hubble} ``\emph{constant}''
 \begin{equation}\label{H}
 H(t)=\frac{\dot{a}}{a},
 \end{equation}
we rewrite this system of equations in the form:
\begin{eqnarray}
\label{Eq_Phi1}
\ddot{\Phi}=&\displaystyle -3H\dot{\Phi}-em^2 \Phi+\alpha\Phi^3;\\
\label{Eq_phi1}
\ddot{\varphi}=&  -3H\dot{\varphi}+\varepsilon \mathfrak{m}^2 \varphi-\beta\varphi^3;\\
\label{dH/dt}
\dot{H}=&\displaystyle -\frac{1}{2}\dot{\Phi}^2+\frac{1}{2}\dot{\varphi}^2.
\end{eqnarray}
However, in this case it is necessary to take into account the Einstein equation for the component $^4_4$ \eqref{EqEinst0_4}, since it is a first-order equation with respect to the scalar potential and the scale factor and limits the arbitrariness of the solutions of the system \eqref{Eq_Phi1} - \eqref{dH/dt}, actually being the first \emph{fixed} integral of this system.

Indeed, let us introduce a quantity useful in what follows - \emph{effective energy of the system}, $\mathcal{E}$:
\begin{eqnarray}\label{E}
\mathcal{E}= \displaystyle \frac{\dot{\Phi}^2}{2}+\frac{e m^2\Phi^2}{2}-\frac{\alpha\Phi^4}{4}\displaystyle -\frac{\dot{\varphi}^2}{2}+\frac{\varepsilon \mathfrak{m}^2\varphi^2}{2}-\frac{\beta\varphi^4}{4}+\Lambda,
\end{eqnarray}
with the help of which the equations \eqref{EqEinst0_4} and the linear combination of the equations \eqref{Eq_Phi1} - \eqref{Eq_phi1}, respectively, can be given the following compact form:
\begin{eqnarray}
\label{EqEinst0_4_E}
3H^2-\mathcal{E}=0,\\
\label{Eq_Phi2}
\dot{\mathcal{E}}+3H(\dot{\Phi}^2-\dot{\varphi}^2)=0.
\end{eqnarray}
Substituting the expression for $ \dot{\Phi} ^ 2- \dot {\varphi} ^ 2 $ from \eqref{dH/dt} into \eqref{Eq_Phi2}, we obtain the conservation law \emph{total energy}
\begin{equation}\label{Einst44=Const}
\frac{d}{dt}(\mathcal{E}-3H^2)=0\Rightarrow E=\mathcal{E}-3H^2 = \mathrm{Const},
\end{equation}
whose partial integral is the Einstein equation \eqref{EqEinst0_4} (or \eqref{EqEinst0_4_E}). Note that in our case, this constant is equal to zero, since the universe is spatially flat. Since $H^2$ is nonnegative, \eqref{EqEinst0_4_E} implies an important property:
\begin{equation}\label{E>=0}
\mathcal{E}\geqslant 0.
\end{equation}
Since, as we noted above, the Einstein equation \eqref{EqEinst0_4}, and in the new notation, the equation \eqref{EqEinst0_4_E} is \emph{the first partial integral of the field equations \eqref{Eq_Phi0} and the Einstein equations \eqref{EqEinst0_4_E}}, hence, the relation \eqref{E>=0} is also a necessary integral condition for the dynamical system \eqref{Eq_Phi0} - \eqref{EqEinst0_1}.

Further, in order to get rid of this arbitrariness, as a basic system of equations, consider the system of field equations \eqref{Eq_Phi0} - \eqref{Eq_phi0} and \eqref{EqEinst0_1}, but in this case, in the equation \eqref{EqEinst0_1} we use the definition of the constant Hubble \eqref{H} and the expression for $\dot{\Phi}^2- \dot{\varphi}^2 $ from \eqref{EqEinst0_4}. Thus, we represent the equation \eqref{EqEinst0_1} in the form:
\begin{eqnarray}\label{EqEinst0}
\dot{H}=\displaystyle -3H^2+\frac{e m^2\Phi^2}{2}-\frac{\alpha\Phi^4}{4}\displaystyle  +\frac{\varepsilon\mathfrak{m}^2\varphi^2}{2}-\frac{\beta\varphi^4}{4}+\Lambda.
\end{eqnarray}
As a consequence of \eqref{Einst44=Const} we have to integrate the indicated system of equations, setting the constant in \eqref{Einst44=Const} equal to zero. This, in turn, means that Einstein's equation \eqref{EqEinst0_4_E} can be used as an initial condition for determining, for example, the Hubble constant:
\begin{equation}\label{H0}
H(t_0)=\pm \sqrt{\frac{\mathcal{E}(t_0)}{3}}.
\end{equation}
In this case, the sign on the right side of \eqref{H0} can be selected according to the problem being solved (the sign ``$ + $'' corresponds to the expansion stage, the sign ``$ - $'' corresponds to the compression stage).

Next, we introduce \emph{invariant cosmological acceleration}
\begin{equation}\label{omega}
\Omega=\frac{a\ddot{a}}{\dot{a}^2}\equiv 1+\frac{\dot{H}}{H^2},
\end{equation}
and represent the equation \eqref{dH/dt} in an equivalent form:
\begin{equation}\label{omega<1}
\Omega=1-\frac{\dot{\Phi}^2}{2H^2}+\frac{\dot{\varphi}^2}{2H^2}.
\end{equation}
Note that the invariant acceleration $\Omega(t)$ is related to the above-introduced barotropic coefficient $w$ by the relation:
\begin{equation}\label{Omega-w}
\Omega=-\frac{1}{2}(1+3w).
\end{equation}
In this article, we will investigate precisely the cosmological acceleration function $ \Omega (t) $, and not the barotropic coefficient $ w(t) $, since the linear relationship between pressure and energy density $ p = w \varrho $ arises \emph{only in private extreme situations}. The more accepted in hydrodynamics relation $ w = dp / d \varrho $, which determines the square of the speed of sound in a medium, is applicable only in those special cases when $ p = p (\varrho) $, which are not realized even in the one-field cosmological model. In addition, in direct astronomical observations, it is the cosmological acceleration that is measured as the difference of the Hubble constants depending on the distance. Note that, according to \eqref{Omega-w}, the values of $w \leqslant-1$ correspond to the values $\Omega \geqslant 1 $, and the values $w \geqslant-1 / 3 $ correspond to the values $\Omega \leqslant0 $. We also note a useful relation for the quadratic invariant of curvature of the Friedmann space \eqref{Freed}, which controls quantum processes:
\begin{eqnarray}\label{sigma}
\sigma\equiv\sqrt{R_{ijkl}R^{ijkl}}=H^2\sqrt{6(1+\Omega^2)}\equiv \sqrt{6\bigl[H^4+\bigl(H^2+\dot{H}\bigr)^2\bigr]}\geq0.
\end{eqnarray}

Finally, we introduce expressions useful for analysis for the energy of the classical, $\mathcal{E}_\Phi $, and phantom, $\mathcal{E}_\varphi$, fields:
\begin{eqnarray}
\label{E_F-E_p}
\mathcal{E}_\Phi=\frac{\dot{\Phi}^2}{2}-\frac{\alpha}{4}\biggl(\Phi^2-\frac{e}{\alpha}\biggr)^2;& \displaystyle
\mathcal{E}_\varphi=-\frac{\dot{\varphi}^2}{2}-\frac{\beta}{4}\biggl(\varphi^2-\frac{\varepsilon}{\beta}\biggr)^2,
\end{eqnarray}
with which the expression \eqref{E} can be made more transparent:
\begin{equation}\label{EFp}
\mathcal{E}=\mathcal{E}_\Phi+\mathcal{E}_\varphi+\Lambda_0.
\end{equation}

\subsection{Normal form of dynamic equations}
Assuming further $ m  \not \equiv0 $ and introducing a dimensionless time variable $\tau$
\begin{equation}\label{tau}
\tau =mt; \quad \biggl(f' =\frac{df}{d\tau}\biggr)
\end{equation}
as well as dimensionless fundamental constants and dimensionless functions
\begin{eqnarray}\label{rename}
\alpha_m=\frac{\alpha}{m^2}; \quad \lambda_m=\frac{\Lambda}{m^2};\quad \lambda^0_m=\frac{\Lambda_0}{m^2};\quad h=\frac{a'}{a}=\frac{H}{m};\quad
\beta_m=\frac{\beta}{m^2};\quad \mu=\frac{\mathfrak{m}}{m},
\end{eqnarray}
rewrite the system of basic equations \eqref{Eq_Phi1}, \eqref{Eq_phi1} and \eqref{dH/dt} in normal form:
\begin{eqnarray}\label{Phi'}
\Phi'=& \displaystyle Z \quad (=P_1); \quad \label{Z'} Z'= \displaystyle  -3hZ-e\Phi+\alpha_m\Phi^3 & (=P_2);\\
\label{z'}
\varphi'=& \displaystyle  z \quad (=P_3); \quad z'=  -3hz+\varepsilon\mu^2\varphi-\beta_m\varphi^3 & (=P_4);\\
\label{h'}
h'= & \displaystyle-3h^2+\frac{e\Phi^2}{2}-\frac{\alpha_m\Phi^4}{4} +\frac{\varepsilon\mu^2\varphi^2}{2}-\frac{\beta_m\varphi^4}{4}+\lambda_m & (=P_5).
\end{eqnarray}

According to \eqref{L->L}, the normalized cosmological constant $\lambda_m $ is related to the value of the normalized bare cosmological constant $ \lambda_m $ as follows:
\begin{equation}\label{l->l}
\lambda^0_m=\lambda_m+\frac{1}{4\alpha_m}+\frac{\mu^4}{4\beta_m}.
\end{equation}
When renormalizing \eqref{tau} - \eqref{rename}, the expressions for the invariant cosmological acceleration \eqref{omega} and \eqref{omega<1} remain invariant with the replacement $ H \to h $ and derivatives $ d / dt \to d / d \tau $, and the expression for the effective energy \eqref{E} is transformed as follows:
\begin{eqnarray}\label{E_m}
\mathcal{E}=&m^2 \mathcal{E}_m; \Rightarrow
\mathcal{E}_m(\Phi,Z,\varphi,z)=&\displaystyle\frac{Z^2}{2}+\frac{e\Phi^2}{2}-\frac{\alpha_m\Phi^4}{4} -\frac{z^2}{2}+
\frac{\varepsilon \mu^2\varphi^2}{2}-\frac{\beta_m\varphi^4}{4}+\lambda_m\geqslant 0.
\end{eqnarray}
We also write in this notation the integral condition \eqref{EqEinst0_4_E} (\eqref{Einst44=Const})
\begin{eqnarray}\label{hE}
3h^2-\frac{Z^2}{2}-\frac{e\Phi^2}{2}+\frac{\alpha_m\Phi^4}{4}
+\frac{z^2}{2}-\frac{\varepsilon \mu^2\varphi^2}{2}+\frac{\beta_m\varphi^4}{4}-\lambda_m=0 \Leftrightarrow E=0.
\end{eqnarray}
In this case, Einstein's equation \eqref {hE} can be used as an initial condition for determining the function $h(\tau_0)$:
\begin{equation}\label{h0}
h(\tau_0)=\pm\sqrt{\frac{\mathcal{E}_m(\tau_0)}{3}}\equiv \mathbf{e}\sqrt{\frac{\mathcal{E}_m(\tau_0)}{3}},
\end{equation}
where $\mathbf{e}=\pm 1$ -- single identifier. Note once again that in the adopted notation all variables and fundamental constants are dimensionless, and also that, due to the integral condition \eqref{hE}, the equation \eqref{h'} is equivalent to the equation
\begin{equation}\label{h'z}
h'=\frac{z^2}{2}-\frac{Z^2}{2}.
\end{equation}
So, the following statement is true:\\
\par
\textbf{Statement 1}. \emph{The investigated cosmological model can be considered as a 5-dimensional dynamical system in the arithmetic phase the space $ \mathbb{R} ^ 5 = \{\Phi, Z, \varphi, z, h \} = \mathbb{R}^3 \bigcup \mathbb{R}^3$, described by the phase trajectory}\\
\begin{equation}\label{r(t)}
\mathbf{r}=\mathbf{r}(\tau);\quad \mathbf{r}=(\Phi,Z,\varphi,z,h)\equiv (x_1,x_2,x_3,x_4,x_5).
\end{equation}
\emph{In this case, the dynamical system} \eqref{Phi'} - \eqref{h'} \emph {is completely determined by an ordered set of 6 dimensionless parameters (fundamental interaction constants)}:
\[\mathbf{P}=[\alpha_m,\beta_m,e,\varepsilon,\mu,\lambda_m].\]
\emph{and an ordered set of 5 initial conditions}:
\[\mathbf{r}_0=[\Phi(0),Z(0),\varphi(0),z(0),\mathbf{e}].\]

In what follows, the three-dimensional phase subspaces $ \mathbb{R}^3 = \{\Phi, Z, h \} $ and $\mathbb{R}^3 = \{\varphi, z, h \} $ will be denoted by the symbols $ \Sigma_\Phi $ and $\Sigma_\varphi $ and call for simplicity the classical and phantom phase subspaces, respectively. Notice, that $\Sigma_\Phi\bigcup \Sigma_\varphi=\mathbb{R}^5$ and $\Sigma_\Phi\bigcap \Sigma_\varphi=\mathbb{R}^1=Oh$.

Note the useful symmetry properties of the \eqref{Phi'} - \eqref{h'} system, namely, its invariance under transformations:
\begin{eqnarray}\label{symm}
\{\tau\to-\tau,\Phi\to\Phi,Z\to-Z,\varphi\to\varphi,z\to-z,h\to-h\}; & \{\tau\to-\tau,\Phi\to-\Phi,Z\to Z,\varphi\to\varphi,z\to-z,h\to-h\};\nonumber\\
\{\tau\to-\tau,\Phi\to\Phi,Z\to-Z,\varphi\to-\varphi,z\to z,h\to-h\}; &
\{\tau\to-\tau,\Phi\to-\Phi,Z\to Z,\varphi\to-\varphi,z\to z,h\to-h\};\nonumber\\
\{\tau\to\tau,\Phi\to\Phi,Z\to Z,\varphi\to-\varphi,z\to-z,h\to h\}; &
\{\tau\to\tau,\Phi\to-\Phi,Z\to -Z,\varphi\to\varphi,z\to z,h\to h\};\nonumber\\
\{\tau\to\tau,\Phi\to-\Phi,Z\to -Z,\varphi\to-\varphi,z\to -z,h\to h\}. &
\end{eqnarray}
These symmetry properties make it possible to simplify the analysis of the dynamical system under study.

\section{Qualitative analysis of a dynamic system}
\subsection{Singular points of the dynamic system}
So, the investigated dynamic system has 5 degrees of freedom, and its state is unequivocally determined by the coordinates of the point $M(\Phi,Z,\varphi,z,h)$  $\equiv M(x_1,x_2,x_3,x_4,x_5)$ in 5-dimensional phase space $\mathbb{R}^5$. Thus the phase space region in which the energy condition is violated \eqref{E_m}, i.e., in which
\begin{equation}\label{E<0}
\Omega\subset \mathbb{R}^3:\quad \mathcal{E}_m(\Phi,Z,\varphi,z)<0,
\end{equation}
are forbidden. As a result, the phase space of the dynamic system can turn out to be multiply connected.  Let us notice that the forbidden regions \eqref{E<0},  if any, are five-dimensional cylinders with axes parallel to $Oh$ and bounded ``above'' and ``below'' by symmetric ``covers'' $h=h_\pm(\Phi,Z,\varphi,z)$, where $h_\pm$ are roots of the equation \eqref{hE}.

Singular points of the dynamical system are determined by equality to zero of the right-hand sides of the normal system of equations. Thus, for the coordinates of singular points of the dynamic system \eqref{Phi'} -- \eqref{h'}  we have the following equations:
\begin{eqnarray}\label{Eq_points_F}
Z=0;\quad  e\Phi-\alpha_m\Phi^3=0;\\
\label{Eq_points_f}
z=0;\quad\varepsilon\mu^2\varphi-\beta_m\varphi^3=0;\\
\label{Eq_points_h}
3h^2=\frac{e\Phi^2}{2}-\frac{\alpha_m\Phi^4}{4} +\frac{\varepsilon\mu^2\varphi^2}{2}-\frac{\beta_m\varphi^4}{4}+\lambda_m.
\end{eqnarray}
Note that according to \eqref{dH/dt} the coordinates of the extremum points $h_{ex}$ of the function $h(t)$ satisfy the condition
\begin{equation}\label{h_min-max}
h_{ex}:\quad Z^2-z^2=0,
\end{equation}
on the other hand according to equation \eqref{h'} the coordinates of these points satisfy the equation \eqref{Eq_points_h}. In the future, the two-dimensional algebraic surface of the fourth order (\ref{Eq_points_h}) will be called the \emph{surface of the effective potential} $\Sigma_{\Phi\varphi h}$. Thus, the following statement is true.\\
\par
\textbf{Statement 2}. \emph{The function $h(\tau)$ minimum and maximum points lie on the surface of the effective potential (\ref{Eq_points_h}), and the singular points of the dynamical system are the extremum points of this surface}  (see Fig. \ref{Surf1}). \\

Coordinates of projections of singular points in two-dimensional planes $\Pi_\Phi=\{\Phi,Z\}$ and $\Pi_\varphi=\{\varphi,z\}$ are determined independently by expressions:
\begin{eqnarray}\label{F0,Z0}
\Phi_0=0;\quad \Phi_\pm=\pm\sqrt{\frac{1}{e\alpha_m}}, \quad (\mbox{if } e\alpha_m>0);\\
\label{f0,z0}
\varphi_0=0;\quad \varphi_\pm =\pm\sqrt{\frac{\mu^2}{\varepsilon\beta_m}}, \quad (\mbox{if } \varepsilon\beta_m>0).
\end{eqnarray}
   The points $M_{0,0}(0,0), M_{\pm1,0}({\Phi_\pm,0})$, $M_{0,\pm1}(0,\varphi_\pm), M_{\pm1,\pm1}(\Phi_\pm,\varphi_\pm)$ on plane $h=0$ define the \emph{base rectangle of the dynamical system}  (see Fig. \ref{base_points}). Moreover, the $h$ -coordinates of these points lie on the symmetric covers $h = h_ \pm (\Phi, 0, \varphi, 0) $. Thus, in the general case, a dynamic system can have 18 singular points\footnote{In what follows, for simplicity of notation, we will omit their zero - coordinates in the coordinates of points $Z,z$.}.
\TwoFigs{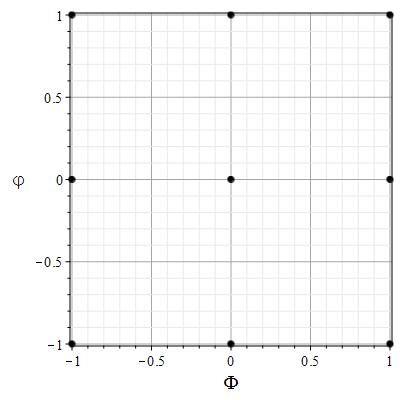}{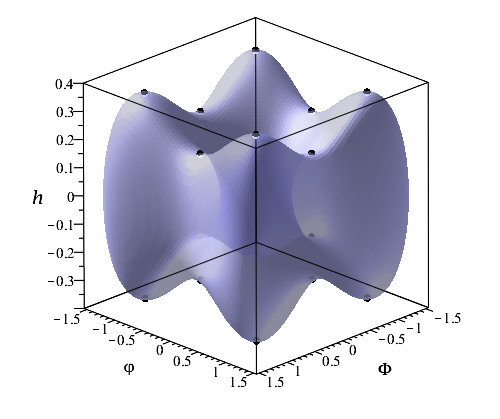}{\label{base_points} Base rectangle of the dynamic system with the parameters $\mathbf{P}=[1,1,1,1,1,-0.1]$. Projections of singular points are shown.}{\label{Surf1}Effective potential surface \eqref{Eq_points_h} $\Sigma_{\Phi\varphi h}$ of the dynamical system with parameters$\mathbf{P}=[1,1,1,1,1,-0.1]$. Projections of singular points are shown.}
These singular points are located at the intersection of the perpendiculars to the 9 points of the base rectangle with the surface of the effective potential \eqref{Eq_points_h}. These are, firstly, two symmetric points located on the $Oh$ --
\begin{equation}\label{M_pm}
M_{0,0}^\pm\biggl(0,0,\pm\sqrt{\frac{\lambda_m}{3}}\biggr);\quad(\mbox{if }\; \lambda_m\geqslant0,\; \forall\alpha_m,\beta_m),
\end{equation}
secondly, 8 points $M^\pm_{\pm1, \pm1} $ symmetric with respect to the plane $ \{\Phi \varphi\}$, located at the intersection of perpendiculars to the vertices of the base rectangle in the plane $\{\Phi, \varphi \} $ and effective potential surfaces, --
\begin{eqnarray}\label{M_1,1}
M^\pm_{\pm1,\pm1}\biggl(\pm\frac{1}{\sqrt{e\alpha_m}},\pm\frac{\mu}{\sqrt{\varepsilon\beta_m}}, \pm \sqrt{\frac{\lambda^0_m}{3}}\biggr),
\quad \biggl(\mathrm{if}\; e\alpha_m>0,\ \varepsilon\beta_m>0,
\lambda^0_m\geqslant0\biggr);
\end{eqnarray}
and,thirdly, 8 points $M^\pm_{0,\pm1}$ and $M^\pm_{\pm1,0}$ symmetric with respect to the plane $\{\Phi,\varphi\}$, located at the intersection of perpendiculars to the midpoints of the sides of the base rectangle in the plane $\{\Phi,\varphi\}$ and effective potential surfaces (see Fig. \ref{Surf1})
\begin{eqnarray}\label{M_0,1}
M^\pm_{0,\pm1}\biggl(0,\pm\frac{\mu}{\sqrt{\varepsilon\beta_m}}, \pm \sqrt{\frac{\lambda_\beta}{3}}\biggr),
& \displaystyle \biggl(\forall\alpha_m,\ \mathrm{if}\; \varepsilon\beta_m>0,\; \lambda_\beta\equiv \lambda_m+\frac{\mu^4}{4\beta_m}\geqslant0\biggr);\\
\label{M_1,0}
M^\pm_{\pm1, 0}\biggl(\pm\frac{1}{\sqrt{e\alpha_m}},0, \pm \sqrt{\frac{\lambda_\alpha}{3}}\biggr),
& \displaystyle \biggl(\mathrm{if}\;e\alpha_m>0,\ \forall\beta_m,\; \lambda_\alpha\equiv \lambda_m+\frac{1}{4\alpha_m}\geqslant0\biggr).
\end{eqnarray}
Note that if we used \eqref{h'} instead of  \eqref{dH/dt}, the
coordinate  $h$ of the singular points would remain arbi-trary, which is precisely a consequence of the above-mentioned arbitrariness.
In total, 7 fundamentally different cases are possible, when the dynamical system has 18,16,12,8,4,2 and 0 singular points. Calculating the value of the effective energy \eqref{E_m} at singular points, we find:
\begin{equation}\label{E_m(M)}
\mathcal{E}_m(M^\pm_{0,0})=\lambda_m;\quad \mathcal{E}_m(M^\pm_{\pm 1,0})=\lambda_\alpha;\quad \mathcal{E}_m(M^\pm_{0,\pm 1})=\lambda_\beta;
\quad\mathcal{E}_m(M^\pm_{\pm 1,\pm10})=\lambda^0_m.
\end{equation}
Thus, from the conditions \eqref{M_pm}, \eqref{M_1,1}, \eqref{M_0,1} and \eqref{M_1,0}  it follows the statement. \\
\par
\textbf{Statement 3}. \emph{All singular points of the dynamic system \eqref{Phi'} -- \eqref{h'}, if they exist, are accessible.}\\
Note further that all singular points of the dynamical system are the extremum points of its effective energy in the plane $\{Z=0,z=0\}$, and therefore are the points of extrema of its effective potential.
Indeed, differentiating $\mathcal{E}_m$ \eqref{E_m}, we find the necessary conditions for a local extremum:
\[\partial_\Phi \mathcal{E}_m=0,\quad \partial_\varphi \mathcal{E}_m=0.\]
Obviously, these conditions coincide with the equations\eqref{Eq_points_F}, \eqref{Eq_points_f}. Calculating the second derivatives, we find
\begin{eqnarray}
a_{11}=\partial^2_{\Phi\Phi}\mathcal{E}_m=e-3\alpha_m\Phi^2;\quad a_{12}=\partial^2_{\Phi\varphi}\mathcal{E}_m=0;\quad a_{22}=\partial^2_{\varphi\varphi}\mathcal{E}_m=\varepsilon\mu^2-3\beta_m\varphi^2.\nonumber
\end{eqnarray}
A sufficient condition for the extremum of the effective energy is the definiteness of the matrix $||a_{ik}||$ -- absolute minimum in case of positive definiteness, and the absolute maximum in the case of negative definiteness. In Fig. \ref{ris_E_m}shows a graph of the effective energy \eqref{E_m} in the plane $\{Z=0,z=0\}$ shows a graph of the effective energy. Note that in this case 4 points located at the midpoints of the sides of the base rectangle correspond to conditional extrema, while the central point corresponds to the absolute minimum, and 4 vertices of the base rectangle correspond to the absolute maximums of the effective energy.

\TwoFigs{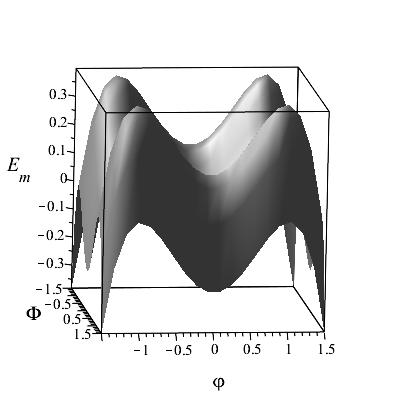}{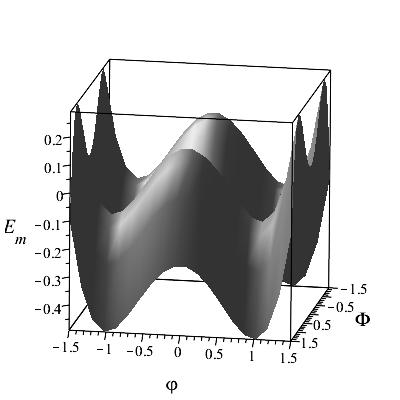}{\label{ris_E_m} Effective energy surface \eqref{E_m} in the plane $\{Z=0,z=0\}$  of a dynamic system with parameters $\mathbf{P}=[1,1,1,1,1,-0.1]$.}{\label{ris_E1_m} Effective energy surface \eqref{E_m} in the plane $\{Z=0,z=0\}$  of a dynamic system with parameters $\mathbf{P}=[-1,1,-1,1,1,-0.1]$.}

In Fig \ref{ris_E1_m} shows a graph effective energy \eqref{E_m} in the plane $\{Z=0,z=0\}$ for a set of parameters different from the one considered above (the signs of the constants in the phantom field potential are reversed). In this case, both the central point and the 4 vertices of the base rectangle correspond to the conditional extrema, 2 points $(\Phi_\pm,0)$ correspond to absolute maxima, and 2 points $(0,\varphi_\pm)$ -- absolute minimum of the effective energy. Note that the base rectangles in both cases coincide.

\subsection{ The character of singular points of the dynamic system\label{character}}
Calculating the matrix of the dynamic system \eqref{Phi'} -- \eqref{h'}  $a_{ik}=\partial_kP_i$, where $P_i$ are the right-hand sides of the dynamic system equations, we find with on account of   \eqref{Eq_points_F} -- \eqref{Eq_points_f}:
{\small
\begin{eqnarray}\label{A}
\mathbf{A}(M)=
\!\!\!\left(\begin{array}{ccccc}
0                   & 1   &   0 & 0 & 0\\
\!\!\!-e\!+\!3\alpha_m\Phi^2  & -3h &   0 & 0 & 0 \\
 0 & 0 & 0 & 1 & 0 \\
 0 & 0 &\!\!\! \varepsilon\mu^2-3\beta_m\varphi^2 & - 3h & 0 \\
0                   & 0    & 0 &  0 & -6h \\
\end{array}\right).\nonumber
\end{eqnarray}
}
Thus, although the dynamical system has 5 degrees of freedom, the matrix of the dynamical system at its singular points has a block-diagonal structure, in which two independent blocks correspond to two components of the scalar doublet, and the fifth dimension $h$ turns out to be independent, which corresponds to an independent eigenvalue matrices $\mathbf{A}(M)$: $k_5=-6h$. The block-diagonal structure of the matrix of the dynamical system, which is a consequence of the absence of direct interaction between the components of the doublet, greatly simplifies the qualitative analysis, reducing it to the analysis of three subsystems, the relationship between which is established by the Einstein equation \eqref{hE}. The determinant of the matrix $\mathbf{A}(M)$ is
\begin{equation}\label{detA}
\mathrm{det}(\mathbf{A})=6h(e-3\alpha_m\Phi^2)(\varepsilon\mu^2-3\beta_m\varphi^2),
\end{equation}
and the characteristic equation for eigenvalues $k$ the matrix \eqref{A}  %
\begin{equation}\label{eigen_number}
\mathrm{det}(\mathbf{A(}M)-k\mathbf{E})=0
\end{equation}
has the form
\begin{eqnarray}
(k+6h)(k^2+3kh+e-3\alpha_m\Phi^2)
(k^2+3kh-\varepsilon\mu^2+3\beta_m\varepsilon^2)=0.
\end{eqnarray}
Thus, the eigenvalues of the matrix $\mathbf{A}(M)$ are
\begin{eqnarray}\label{k_i}
k_{1,2}=&\displaystyle-\frac{3}{2}h\pm \frac{1}{2}\sqrt{9h^2-4e+12\alpha_m\Phi^2};&\nonumber\\[6pt]
k_{3,4}=&\displaystyle-\frac{3}{2}h\pm \frac{1}{2}\sqrt{9h^2+4\varepsilon\mu^2-12\beta_m\varphi^2};&\nonumber\\[6pt]
k_5=&-6h.&\nonumber
\end{eqnarray}
In expressions for the eigenvalues, it is also necessary to use the Einstein equation \eqref{hE} to calculate $h^2$.
At singular points, the eigenvalues of the matrix of the dynamical system are
$$
\begin{array}{ll}
$$
\begin{array}{l}
M_{0,0}^\pm:\quad k_5=\mp  2\sqrt{3\lambda_m}, \quad \lambda_m\geq 0 \\[6pt]
k_{1,2}=\mp\dfrac{1}{2}\sqrt{3\lambda_m}\pm \dfrac{1}{2}\sqrt{3\lambda_m-4e};\\[6pt]
k_{3,4}=\mp\dfrac{1}{2}\sqrt{3\lambda_m}\pm \dfrac{1}{2}\sqrt{3\lambda_m+4\varepsilon\mu^2};
\end{array}
$$&
$$
\begin{array}{l}
M_{\pm 1,\pm 1}^\pm:\quad k_5=\mp  2\sqrt{3\lambda^0_m}, \quad \lambda^0_m\geq 0 \\[6pt]
k_{1,2}=\mp\dfrac{1}{2}\sqrt{3\lambda^0_m}\pm \dfrac{1}{2}\sqrt{3\lambda^0_m+8e};\\[6pt]
k_{3,4}=\mp\dfrac{1}{2}\sqrt{3\lambda^0_m}\pm \dfrac{1}{2}\sqrt{3\lambda^0_m-8\varepsilon\mu^2};
\end{array}
$$
\vspace{18pt}
\\
$$
\begin{array}{l}
M_{0,\pm 1}^\pm:\quad k_5=\mp  2\sqrt{3\lambda_{\beta}}, \quad \lambda_{\beta}\geq 0 \\[6pt]
k_{1,2}=\mp\dfrac{1}{2}\sqrt{3\lambda_{\beta}}\pm \dfrac{1}{2}\sqrt{3\lambda_{\beta}-4e};\\[6pt]
k_{3,4}=\mp\dfrac{1}{2}\sqrt{3\lambda_{\beta}}\pm \dfrac{1}{2}\sqrt{3\lambda_{\beta}-8\varepsilon\mu^2};
\end{array}
$$&
$$
\begin{array}{l}
M_{\pm 1,0}^\pm:\quad k_5=\mp  2\sqrt{3\lambda_{\alpha}}, \quad \lambda_{\alpha}\geq 0 \\[6pt]
k_{1,2}=\mp\dfrac{1}{2}\sqrt{3\lambda_{\alpha}}\pm \dfrac{1}{2}\sqrt{3\lambda_{\alpha}+8e};\\[6pt]
k_{3,4}=\mp\dfrac{1}{2}\sqrt{3\lambda_{\alpha}}\pm \dfrac{1}{2}\sqrt{3\lambda_{\alpha}+4\varepsilon\mu^2};
\end{array}
$$
\end{array}
$$
In pairs of symmetric points, the eigenvalues of the matrix of the dynamical system coincide. The signs in front of the first and second radicals in these formulas take on independent meanings: the signs in front of the first radicals correspond to different pairs of dots, the signs in front of the second radicals correspond to different eigenvalues. It is important to remember the necessary conditions for the existence of singular points \eqref{M_pm}, \eqref{M_1,1}, \eqref{M_0,1} and \eqref{M_1,0}. Going through all possible options, we come to the following result (Tab. \ref{Tab1}, Тab. \ref{Tab2}).\\[6pt]

{\small
\begin{minipage}{.4\textwidth}
\refstepcounter{table} Table \thetable. \label{Tab1} The character of singular points\newline of the dynamic system for  classic field\\

\begin{center}
\begin{tabular}{|l|r|l|}
\hline
Singular points&$e$&Point type\\
\hline
\multirow{2}{*}{$M_{0,0}^+$}&1&Attractive\\
&-1&Saddle\\
\hline
\multirow{2}{*}{$M_{0,0}^-$}&1&Repulsive\\
&-1&Saddle\\
\hline
\multirow{2}{*}{$M_{\pm 1,\pm 1}^+$}&1&Saddle\\
&-1&Attractive\\
\hline
\multirow{2}{*}{$M_{\pm 1,\pm 1}^-$}&1&Saddle\\
&-1&Repulsive\\
\hline
\multirow{2}{*}{$M_{0,\pm 1}^+$}&1&Attractive\\
&-1&Saddle\\
\hline
\multirow{2}{*}{$M_{0,\pm 1}^-$}&1&Repulsive\\
&-1&Saddle\\
\hline
\multirow{2}{*}{$M_{\pm 1,0}^+$}&1&Saddle\\
&-1&Attractive\\
\hline
\multirow{2}{*}{$M_{\pm 1,0}^-$}&1&Saddle\\
&-1&Repulsive\\
\hline
\end{tabular}
\end{center}
\end{minipage}
\begin{minipage}{.4\textwidth}
\refstepcounter{table}\label{Tab2}Table \thetable. The character of singular points \newline of the dynamic system for phantom field\\
\begin{flushright}
\begin{tabular}{|l|r|l|}
\hline
Singular points&$\varepsilon$&Point type\\
\hline
\multirow{2}{*}{$M_{0,0}^+$}&1&Saddle\\
&-1&Attractive\\
\hline
\multirow{2}{*}{$M_{0,0}^-$}&1&Saddle\\
&-1&Repulsive\\
\hline
\multirow{2}{*}{$M_{\pm 1,\pm 1}^+$}&1&Attractive\\
&-1&Saddle\\
\hline
\multirow{2}{*}{$M_{\pm 1,\pm 1}^-$}&1&Repulsive\\
&-1&Saddle\\
\hline
\multirow{2}{*}{$M_{0,\pm 1}^+$}&1&Attractive\\
&-1&Saddle\\
\hline
\multirow{2}{*}{$M_{0,\pm 1}^-$}&1&Repulsive\\
&-1&Saddle\\
\hline
\multirow{2}{*}{$M_{\pm 1,0}^+$}&1&Saddle\\
&-1&Attractive\\
\hline
\multirow{2}{*}{$M_{\pm 1,0}^-$}&1&Saddle\\
&-1&Repulsive\\
\hline
\end{tabular}
\end{flushright}
\end{minipage}
}

\vspace{6pt}
In this case, the attracting singular points correspond to the negative signs of the real parts of the eigenvalues, the repulsive ones are positive, the saddle points are opposite. The character of the singular points in the upper ($h>0$) and lower ($h<0$) can be connected using the symmetry relations \eqref{symm} taking into account the fact that the change in the sign of the time $\tau\to-\tau$ corresponds to a change in the character of the singular point to the opposite (attraction $\leftrightarrow$ repulsion), while the saddle character of the points remains invariant.\\[4pt]
\subsection{Exact solutions in attractive singular points}
Attractive singular points correspond to stable exact solutions.
The corresponding exact solutions are determined by the constant coordinates of these points  $\Phi=\Phi_0,Z=0,\varphi=\varphi_0,z=0$ according to \eqref{M_pm} -- \eqref{M_1,0}, and in these points
\begin{equation}\label{h_0}
h_0=\pm\sqrt{\frac{\lambda^0_m}{3}}.
\end{equation}
These solutions correspond to expansion with constant speed $H=\mathrm{Const}>0$ ($\Omega=+1$) for $h_0>0$ or compression with constant speed $H=\mathrm{Const}<0$ ($\Omega=+1$) for $h_0<0$. In this case, stable exact solutions will correspond to attractive singular points, and unstable exact solutions to repulsive and saddle singular points. Relative scale factor of these exact solutions are
\begin{equation}\label{sol_inf}
a(\tau)=a_0 \mathrm{e}^{h_0\tau}, \quad (\tau\in(-\infty,+\infty)).
\end{equation}
Going through all possible cases in the tables  \ref{Tab1} and \ref{Tab2}, indicate \emph{stable exact solutions}:
\begin{enumerate}
\item $M^+_{0,0}$ for $e=+1,\alpha>0;\ \varepsilon=-1,\beta<0$;
\item $M^+_{\pm1,\pm1}$ for $e=-1,\alpha<0;\ \varepsilon=+1,\beta>0$;
\item $M^+_{0,\pm1}$ for $e=+1,\alpha>0;\ \varepsilon=+1,\beta>0$;
\item $M^+_{\pm1,0}$ for $e=-1,\alpha<0;\ \varepsilon=-1,\beta<0$.
\end{enumerate}

Note that if we select from these cases only those that correspond to \emph{pure Higgs fields} ($\alpha>0,\beta>0,e=+1,\varepsilon=+1$), then from this list only one remains the 3rd case: $M^+_{0,\pm1}$. Further, in all the indicated cases the stationary solution for the  \emph{nontrivial scalar doublet} $\{\Phi\not=0,\varphi\not=0\}$ exists only in the second case, but this case does not correspond to the pure Higgs potential, because $e=-1,\alpha<0$. Thus, the following statement is true.\\
\par
\textbf{Statement 4}. \emph{In total, there are 4 classes of exact stable stationary solutions of dynamic equations \eqref{Phi'} -- \eqref{h'}
with the integral condition \eqref{hE}, at that, in the case of purely Higgs potentials for a nontrivial scalar doublet in which both scalar fields are present, there are no exact stable stationary solutions. Wherein, there are stable stationary solutions for scalar Higgs singlets $\varphi\equiv0$ or $\Phi\equiv 0$, this is $M^+_{\pm1,0}$ or $M^+_{0,\pm1}$ respectively, and also purely vacuum stationary solutions $M^+_{0,0}$.} \\

This is a very important conclusion that has obvious cosmological applications for the asymmetric scalar doublet model. Note that this statement does not exclude the possibility of an infinite approximation of trajectories to solutions $h = h_0$, i.e., to asymptotically inflationary solutions.

\subsection{Stable solutions with a constant field of one of the components of the doublet}
If one of the fields of the doublet coincides with the coordinates of a stable singular nonzero point in the corresponding subspace, the dimension of the dynamical system \eqref{Phi'} - \eqref{h'} is reduced by two, i.e. the dynamical system becomes three-dimensional. Let us study this possibility.  There can be only two such cases (taking into account the symmetry of the system):
\begin{eqnarray}
\label{Phi0}
\mathbf{1.}\; \Phi=\Phi_\pm,\; Z=0 \quad (e\alpha_m>0),\; \varphi=\varphi(\tau);\\
\label{varphi0}
\mathbf{2.}\;\; \varphi=\varphi_\pm,\;\; z=0 \quad (\varepsilon\beta_m>0),\; \Phi=\Phi(\tau)
\end{eqnarray}
(see \eqref{F0,Z0} -- \eqref{f0,z0}).
In the case of \textbf{1} the system of dynamic equations is reduced to the following:
\begin{eqnarray}
\label{f'1-z'1}
\varphi'=& z \displaystyle; \quad z'= -3hz+\varepsilon\mu^2\varphi-\beta_m\varphi^3; & \\
\label{h'f}
h'=& \displaystyle-3h^2-\frac{\beta_m}{4}\biggl(\varphi^2-\frac{\mu^2}{\varepsilon\beta_m}\biggr)^2+\lambda^0_m &
\end{eqnarray}
with the integral condition
\begin{eqnarray}\label{hEf1}
3h^2+\frac{z^2}{2}+\frac{\beta_m}{4}\biggl(\varphi^2-\frac{\mu^2}{\varepsilon\beta_m}\biggr)^2-\lambda^0_m=0,
\end{eqnarray}
and in the case of \textbf{2} is reduced to the following:
\begin{eqnarray}
\label{F'1-Z'1}
\Phi'=&\displaystyle Z;\quad Z'= -3hZ-e\Phi+\alpha_m\Phi^3 ; & \\
\label{h'F}
h'=& \displaystyle-3h^2-\frac{\alpha_m}{4}\biggl(\Phi^2-\frac{1}{e\alpha_m}\biggr)^2+\lambda^0_m &
\end{eqnarray}
with the integral condition
\begin{eqnarray}\label{hEF1}
3h^2-\frac{Z^2}{2}+\frac{\alpha_m}{4}\biggl(\Phi^2-\frac{1}{e\alpha_m}\biggr)^2-\lambda^0_m=0.
\end{eqnarray}

In the appropriate cases of integral conditions \eqref{hEf1} or \eqref{hEF1} equations for  $h(\tau)$ \eqref{h'f} or \eqref{h'F} can also be written in the form:
\begin{eqnarray}
\label{h'f1}
\eqref{h'f}\ \Leftrightarrow\ h'=\frac{z^2}{2};\\
\label{h'F1}
\eqref{h'F}\Leftrightarrow h'=-\frac{Z^2}{2}.
\end{eqnarray}

Obviously, if the solution to one of the above three-dimensional dynamical systems is stable, then the corresponding solution in the complete phase space  $\mathbb{R}^5$ will also be stable. Thus, the following statement is true:\\
\par
\textbf{Statement 5}.\emph{ There are subclasses of stable solutions in which one of the scalar fields is constant and corresponds to a singular point in the corresponding subspace (\eqref{Phi0} or \eqref{varphi0}), while the second field of the doublet is determined by the three-dimensional dynamical system of the corresponding scalar singlet with the cosmological constant $\lambda^0_m$.}\\

\subsection{Asymptotic behavior of phase trajectories}
Due to the block-diagonal structure of the dynamic system matrix  $\mathbf{A}(M)$, its eigenvalues are determined by the characteristic equations in the corresponding two-dimensional planes, and the eigenvectors $\mathbf{u}_{k}^{(M)}$,  corresponding to these eigenvalues
\begin{equation}\label{eigen_vector}
\bigl(\mathbf{A}(M)-k_{i} \mathbf{E}\bigr) \mathbf{u}^{(M)}_{i}=0;
\end{equation}
located in pairs of different phase planes
\[\left\{\mathbf{u}_{1}^{(M)}, \mathbf{u}_{2}^{(M)}\right\} \subset \Sigma_{\Phi}, \quad\left\{\mathbf{u}_{3}^{(M)}, \mathbf{u}_{4}^{(M)}\right\} \subset \Sigma_{\varphi}\ .\]
This fact makes it possible to significantly simplify the qualitative analysis of phase trajectories near the singular point $ M^\alpha $ and reduce it to considering combinations of characteristics of the dynamical system in three-dimensional planes $\Sigma_{\Phi}$, $\Sigma_{\varphi}$.

According to the qualitative theory of differential equations (see, for example, \cite{Bogoyav}) \emph{the asymptotic expression} for the radius vector of the phase trajectory in the neighborhood of the singular point is
 \begin{equation}\label{asymptot1}
 \mathbf{r}(\tau) =\mathbf{r}^{(\alpha)}+\mathbf{Re}\left(\sum_{j=1}^{n} C_{j} \mathbf{u}_{j}^{(\alpha)} \mathrm{e}^{i k_{j}^{(\alpha)} \tau}\right),            \end{equation}
where $\mathbf{r}^\alpha$ is radius vector of the singular point $M^\alpha$. To determine the asymptotic trajectory, it is necessary to determine the constants $C_j$ on the right-hand side of \eqref{asymptot1}. This can be done in a variety of ways, each of which gives a better approximation of the trajectory in different cases. The simplest way is as follows. Substituting the coordinates of the starting point $\mathbf{r}_0$ into the left side \eqref{asymptot1}, we get the equality:
\begin{equation}\label{C-1}
\mathbf{Re}\left(\sum_{j=1}^{n} C_{j} \mathbf{u}_{j}^{(\alpha)} \right)=\mathbf{r}_0-\mathbf{r}^{(\alpha)}.
\end{equation}

\TwoFigs{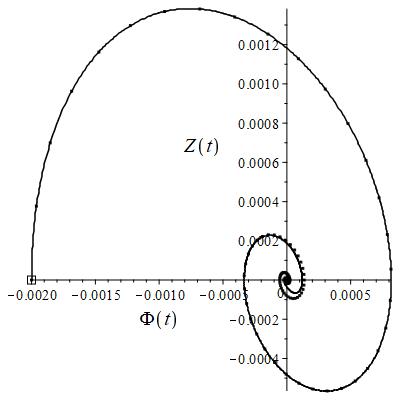}{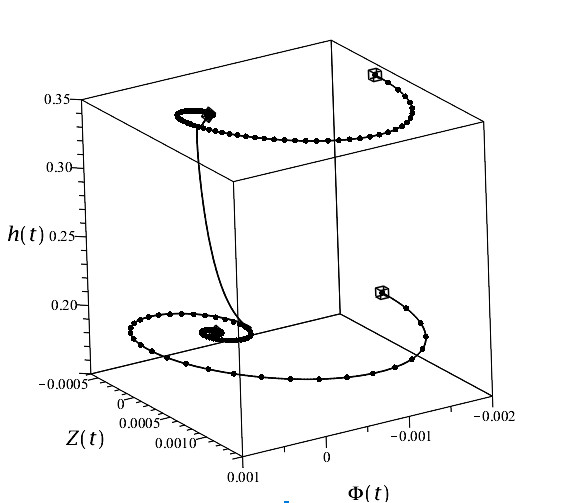}{\label{Asipm_Phi_z} Projection of the phase trajectory in the plane $\{\Phi,Z\}$ (solid line) and  asymptote (dotted line) for parameters $\mathbf{P}=[1,1,1,1,1,0.1]$ and initial conditions $\mathbf{r}_0=[-0.002,0,0,0,1]$. Solid circle shows the projection of the nearest singular point, square shows the projection of the starting point.}{\label{Asipm_Phi_z_h_2} The projection of the phase trajectories in the subspace $\Sigma_{\Phi}$ (solid line) and  asymptote (dotted line) for parameters   $\mathbf{P}=[1,1,1,1,1,0.1]$ and initial conditions   $\mathbf{r}_0=[-0.002,0,0.001,0,-1]$ (bottom line) and $\mathbf{r}_0=[-0.002,0,-1,0,1]$ (top line). Spheres shows the projection of the nearest singular point, cube shows the projection of the starting points.}

The relation \eqref{C-1} can be considered as a system of equations for the unknown constants $C_j$. This method gives good results near the singular point $M^\alpha$. Fig. \eqref{Asipm_Phi_z} and Fig. \eqref{Asipm_Phi_z_h_2} shows examples of constructing asymptotic trajectories using this method. It can be seen that the method gives good results near singular points, but does not track a sharp change in the trajectory near two singular points at once, which is demonstrated in Fig. \ref{Asipm_Phi_z_h_2}. Consider another way to define the constants $C_j$. Differentiating the equation of the asymptotic trajectory \eqref{asymptot1} at the initial point, we obtain the system of equations:
\begin{equation}\label{C-2}
\mathbf{Re}\left(\sum_{j=1}^{n} C_{j}k_j \mathbf{u}_{j}^{(\alpha)} \right)=\mathbf{\dot{r}}_0.
\end{equation}
Substituting the initial conditions $\mathbf{r}_0$ into the right-hand side of the dynamical equations \eqref{Phi'} -- \eqref{h'}, we will find tangent vector
$\mathbf{\dot{r}}_0$ and thus, using the equations \eqref{C-2} we find  the unknown coefficients $C_j$. This method allows to find the phase trajectories and away from the singular points. The combination of these two methods makes it possible to determine asymptotic trajectories quite well.

\section{Einstein-Higgs hypersurface}

Consider the surface \eqref{hE} \emph{total energy} $E=0$ (see \eqref{Einst44=Const})
\begin{eqnarray}\label{Sigma_E}
\Sigma_E:\quad 3h^2-\frac{Z^2}{2}-\frac{\alpha_m}{4}\biggl(\Phi^2-\frac{e}{\alpha_m}\biggr)^2
+\frac{z^2}{2}-\frac{\beta_m}{4}\biggl(\varphi^2-\frac{\varepsilon\mu^2}{\beta_m}\biggr)^2-\lambda^0_m=0,
\end{eqnarray}
which in the future we will call the \emph{Einstein – Higgs hypersurface}. The topological properties of this hypersurface are very important for the analysis of the phase trajectories of the cosmological model, since all phase trajectories, like the singular points of the dynamical system, must lie on this hypersurface.

From the equation \eqref{h'z}, and equation \eqref{h'f1} and \eqref{h'F1}, it follows that for the \emph{classical scalar singlet} $\varphi\equiv0$ or $\varphi=\varphi_\pm$ the function $h(\tau)$ is non-increasing:
\begin{equation}\label{h'F1<0}
h'\leqslant0; \quad (\varphi=0;\varphi_\pm),
\end{equation}
and for \emph{phantom scalar singlet} $\Phi\equiv0$ or  $\Phi=\Phi_\pm$ the function $h(\tau)$ is non-decreasing:
\begin{equation}\label{h'f1>0}
h'\geqslant0; \quad (\Phi=0;\Phi_\pm).
\end{equation}
In the case of a complete scalar doublet, the behavior of the function $h(\tau)$ is the result of the summary action of the components of the scalar doublet, interacting via gravity through the Einstein equation \eqref{Sigma_E}. Therefore, the possibilities for the transition of the cosmological model from the expansion mode to the contraction mode and backward are determined by the topology of the Einstein - Higgs hypersurface \eqref{Sigma_E}, which allows or forbids phase trajectories of the dynamical system \eqref{Phi'} -- \eqref{h'}  to connect points on the Einstein - Higgs hypersurface with positive and negative values $h$, corresponding to the transitions  $h_+\to h_-$ or $h_-\to h_+$. In this regard, it becomes extremely important to study the properties of the Einstein - Higgs hypersurface \eqref{Sigma_E} depending on the values of the fundamental parameters $\mathbf{P}$ of the scalar doublet.

\subsection{Evolution of three-dimensional sections of the Einstein - Higgs\\ hypersurface along phase trajectories}

We investigate the properties of the projections of this hypersurface in each of three-dimensional subspaces
$\Sigma_\Phi$ and $\Sigma_\varphi$ of the phase space, assuming the remaining coordinates are fixed, $\{\varphi_0,z_0\}$ and $\{\Phi_0,Z_0\}$, respectively.

{\note Three-dimensional sections of the Einstein - Higgs hypersurface in each of the subspaces $\Sigma_{\Phi}$ and $\Sigma_{\varphi}$ depend on pairs of dynamical variables of the other field, when changing the sectional shape of which may vary considerably. Therefore, at each moment of time that determines the  phase coordinates of the dynamical system, the three-dimensional sections of the Einstein - Higgs hypersurface will also change.}\\

In Fig. \ref{evol_clas_surf} and \ref{evol_fant_surf} shows the evolution of the classical and phantom projection of the Einstein - Higgs hypersurface for the particular case of a phase trajectory. These figures show how the process of surface merging in the subspace $\Sigma_\Phi$ occurs along the $OZ$ axis while the surface is disrupted simultaneously in the subspace $\Sigma_\varphi$ along the $Oz$ axis.
\begin{figure}[h]
\includegraphics[width=.9\textwidth]{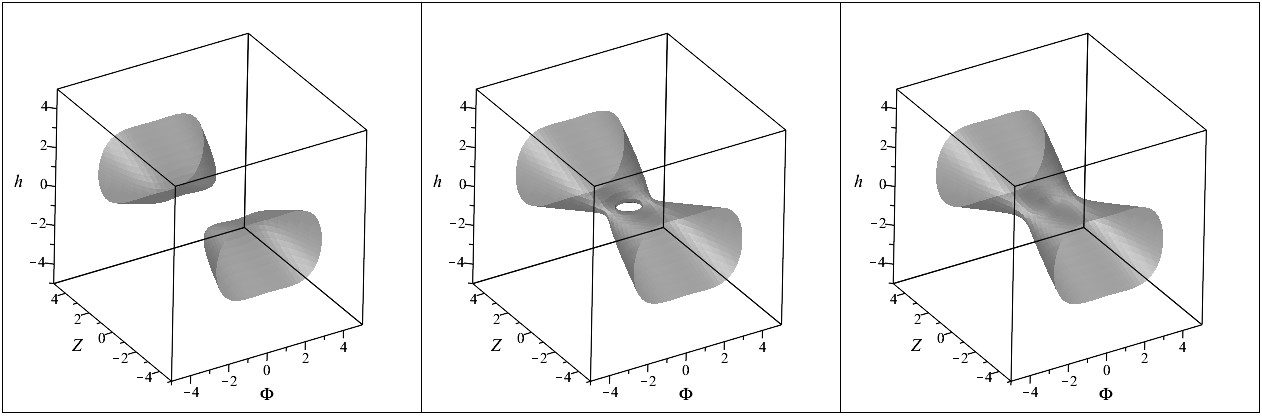}
\caption{\label{evol_clas_surf} Evolution of the projection of the Einstein-Higgs hypersurface in the subspace $\Sigma_\Phi$ for parameters $\mathbf{P}=[1,1,1,1,1,-0.1]$ and initial conditions $\mathbf{r}_0=[0.1,1,1,1,1]$; $\tau=-1,1,10$.}
\end{figure}

\begin{figure}[h]
\includegraphics[width=.9\textwidth]{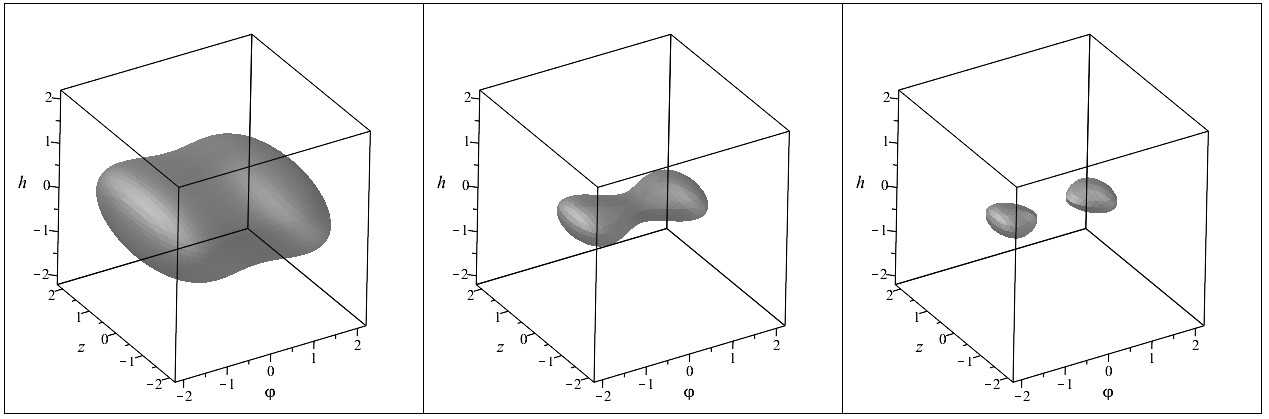}
\caption{\label{evol_fant_surf} Evolution of the projection of the Einstein-Higgs hypersurface in the subspace $\Sigma_\varphi$ for parameters $\mathbf{P}=[1,1,1,1,1,-0.1]$ and initial conditions $\mathbf{r}_0=[0.1,1,1,1,1]$; $\tau=-1,1,10$.}
\end{figure}


\subsection{Classification of three-dimensional sections of the Einstein - \\Higgs hypersurface}
We investigate the three-dimensional projection of the fourth order Einstein - Higgs hypersurface   \eqref{hE} in the subspaces $\Sigma_\Phi$ and $\Sigma_\varphi$. Noting that this hypersurface described biquadratic equation, we make the change of variables
\begin{equation}\label{tilde_F,f}
\tilde{\Phi}=\Phi^2-\displaystyle\frac{1}{e\alpha_m}, \quad \tilde{\varphi}=\varphi^2-\displaystyle\frac{\mu^2}{\varepsilon\beta_m},
\end{equation}
equal to zero at singular points of the dynamical system \eqref{F0,Z0} and \eqref{f0,z0} ($\Phi\not=0,\ \varphi\not= 0$) and reduce the equations of these surfaces to the form
\begin{eqnarray}\label{Sigma_1}
\Sigma_\Phi:& 3h^2-\displaystyle\frac{Z^2}{2}+\displaystyle\frac{\alpha_m\tilde{\Phi}^2}{4}=\mathrm{C}_\Phi,\\
\label{Sigma_2}
\Sigma_\varphi:& 3h^2+\displaystyle\frac{z^2}{2}+\displaystyle\frac{\beta_m\tilde{\varphi}^2}{4}=\mathrm{C}_\varphi,
\end{eqnarray}
where
\begin{eqnarray}\label{CC}
\mathrm{C}_\Phi=-\frac{z_0^2}{2}-\frac{\beta_m\tilde{\varphi_0}^2}{4}+\lambda^0_m\equiv \mathcal{E}^m_\varphi+\lambda^0_m,\\
\mathrm{C}_\varphi=\frac{Z_0^2}{2}-\frac{\alpha_m\tilde{\Phi_0}^2}{4}+\lambda^0_m\equiv \mathcal{E}^m_\Phi+\lambda^0_m,
\end{eqnarray}
 $\mathcal{E}^m_\Phi=\mathcal{E}_\Phi$ and $\mathcal{E}^m_\varphi=\mathcal{E}_\varphi$ are the reduced energies of the classical and phantom fields \eqref{E_F-E_p}. Thus, the introduced ``constants'' $C_\Phi$  $C_\varphi$, in fact, are \emph{effective energy of the phantom and classical fields}, respectively. In the variables $\tilde{\Phi},Z,\tilde{\varphi},z$ the surfaces \eqref{Sigma_1} and \eqref{Sigma_2} or a given additional pair of variables are second-order central surfaces. Since a pair of additional variables changes during the evolution of a dynamical system, the values \eqref{CC} are functions of time: $\mathrm{C}_\Phi(\tau),\ \mathrm{C}_\varphi(\tau)$, which leads to the above-mentioned evolution of the projections of the Einstein - Higgs hypersurface. Further, the transformation \eqref{tilde_F,f} is not bijective, therefore, passing to the initial variables, we obtain pairs of solutions $\Phi=\pm \sqrt{\tilde{\Phi}+1/e\alpha_m}$; $\varphi=\pm \sqrt{\tilde{\varphi}+\mu^2/\varepsilon\beta_m}$.
As a result, this leads, firstly, to deformation of the second-order canonical surfaces (in particular, to their duplication), and, secondly, to the appearance of gaps in the regions
\[\tilde{\Phi}+\frac{1}{e\alpha_m}<0;\quad \tilde{\varphi}+\frac{\mu^2}{\varepsilon\beta_m}<0. \]
At potential infinity $\Phi\to\pm\infty$ or $\varphi\to\pm\infty$ in each of the three-dimensional subspaces depending on the parameters $\{\alpha_m,\beta_m,\mathrm{C}_\Phi(\tau)$, $\mathrm{C}_\varphi(\tau) \}$  these surfaces are similar to the following convex surfaces: either a one-/two-sheet hyperboloid, or an ellipsoid, or a cone (see Tab. \ref{Tab3}, \ref{Tab4}). However, near the singular points, the convexity of the surfaces can be violated.

Below, using the terms ``hyperboloid'' and ``ellipsoid'', etc., everywhere we mean ``deformed hyperboloid'', ``deformed ellipsoid'', etc.

\begin{center}
\refstepcounter{table}
Table \thetable.\label{Tab3}
Einstein-Higgs hypersurface projection type\\
in three-dimensional subspacе $\Sigma_\Phi$ of phase space\\[6pt]

\begin{tabular}{|l|l|l|}
\hline
№&Parameters&Surface type\\
\hline
1.&$\alpha_m\mathrm{C}_\Phi>0$&One-sheet hyperboloid\\
\hline
2.&$\alpha_m\mathrm{C}_\Phi<0$&Two-sheet hyperboloid\\
\hline
3.&$\mathrm{C}_\Phi=0$&Cone second order\\
\hline
\end{tabular}
\end{center}

\begin{center}
\refstepcounter{table}
Table \thetable. \label{Tab4} Einstein-Higgs hypersurface projection type\\
in three-dimensional subspacе $\Sigma_\varphi$ of phase space\\[6pt]

\begin{tabular}{|l|l|l|}
\hline
№&Parameters&Surface type\\
\hline
1.&$\beta_m<0,\mathrm{C}_\varphi>0$&One-sheet hyperboloid\\
\hline
2.&$\beta_m<0,\mathrm{C}_\varphi<0$&Two-sheet hyperboloid\\
\hline
3.&$\beta_m<0, \mathrm{C}_\varphi=0$&Cone second order\\
\hline
4.&$\beta_m>0, \mathrm{C}_\varphi>0$&Ellipsoid\\
\hline
5.&$\beta_m>0, \mathrm{C}_\varphi<0$&Imaginary ellipsoid\\
\hline
6.&$\beta_m>0, \mathrm{C}_\varphi=0$&Imaginary cone\\
\hline
\end{tabular}
\end{center}

Let us analyze the shape of the orthogonal section $\mathbf{S}$ of the Einstein-Higgs hypersurface in the plane perpendicular to the principal axis and passing through its origin. In this case, the principal axis of the Einstein hypersurface is determined by that variable, the sign of the coefficient for which is not repeated in the equations \eqref{Sigma_1} or \eqref{Sigma_2}. Thus, in the case of a classical scalar field, the principal axis is determined by the sign of $\alpha_m$, and in the case of a phantom scalar field, the principal axis is determined by the sign of $\beta_m$. For a classical field with $\alpha_m>0$ the principal axis is $OZ$, otherwise -- $Oh$. For a phantom field with $\beta_m<0$ the principal axis is $O\varphi$, otherwise, we have an ellipsoid (possibly imaginary).
In what follows, $\mathbf{E}$ denotes the ellipse, $A,B$ denotes the points, $\mbox{\large\bf $\infty$}$ denotes the figure eight (see Tab \ref{Tab5} and \ref{Tab6}).\\[6pt]
\begin{tabular}{lr}
\refstepcounter{table}\label{Tab5}
\parbox{7cm}{Table \thetable. Section $\mathbf{S}$ in subspace $\Sigma_\Phi$\\[12pt]
\begin{tabular}{l|l}
$I_\Phi$ & $\alpha_m C_{\Phi}<0\rightarrow \mathbf{S}=\emptyset $;\\
$II_\Phi$ & $\alpha_m C_{\Phi}=0\rightarrow \mathbf{S}=\{A,B\}$;\\
$III_\Phi$ & $0<\alpha_m C_{\Phi}<1/4\rightarrow \mathbf{S}=\{\mathbf{E}_a,\mathbf{E}_b\}$;\\
$IV_\Phi$ & $\alpha_m C_{\Phi}=1/4 \rightarrow \mathbf{S}=\mbox{\large\bf $\infty$}$ ;\\
$V_\Phi$ & $\alpha_m C_{\Phi}>1/4 \rightarrow \mathbf{S}=\mathbf{E}$.\\
\end{tabular}} &
\parbox{7cm}{\refstepcounter{table}\label{Tab6}
Table \thetable. Section $\mathbf{S}$ in subspace $\Sigma_\varphi$\\[12pt]
\begin{tabular}{l|l}
$I_\varphi$ & $C_{\varphi}<0\rightarrow \mathbf{S}=\emptyset$; \\
$II_\varphi$ & $C_{\varphi}=0\rightarrow \mathbf{S}=O(0,0)$;\\
$III_\varphi$ & $C_{\varphi}>0 \rightarrow \mathbf{S}=\mathbf{E}$.\\
\end{tabular}}\\
\end{tabular}

\subsection{Graphic illustrations of the types of sections of the Einstein - Higgs hypersurface}
Note that, according to Table \ref{Tab4}, in the case of a pure Higgs phantom field ($\varepsilon=+1,\beta>0$), the projection of the Einstein-Higgs hypersurface in the phantom subspace $\Sigma_\varphi$ exists only under the condition $C_\varphi>0$ and in this case is a deformed ellipsoid. Therefore, taking into account the definitions \eqref{M_1,1}, \eqref{tilde_F,f} and \eqref{Sigma_2} real phase trajectories of an asymmetric scalar doublet in the case of a pure Higgs phantom field can exist only in the range of values $D(\Phi,Z)$:
\begin{equation}\label{D}
D(\Phi,Z):\quad \mathrm{C}_\varphi>0 \Leftrightarrow
\mathcal{E}^m_\Phi+\lambda^0_m>0.
\end{equation}
Thus, the following statement is true.\\
\par
\textbf{Statement 6}. \emph{Real solutions of dynamical equations \eqref{Phi'} -- \eqref{h'} with integral condition \eqref{hE} in the case of a pure Higgs phantom field ($\varepsilon=+1,\beta>0$) exist only in the domain positive values of the effective energy of the classical scalar field \eqref{D}. Moreover, the section $\mathbf{S}$ in the phantom subspace is a deformed ellipse.}\\

Below, in Fig. \ref{Sect0F3} -- \ref{Sect1p3} the graphs of the listed types of sections of the Einstein - Higgs hypersurface are shown.

\subsubsection*{Type $\mathbf{I}_\Phi$: doubly connected Einstein-Higgs hypersurface: $\Sigma_\Phi$ $\mathbf{P}=[-1,1,-1,1,1,1]$;\\ $\mathbf{I}=[0.2,0.2,0.2,0.2,\pm1]$:  $\alpha_mC_{\Phi}<0$)}

\TwoFigs{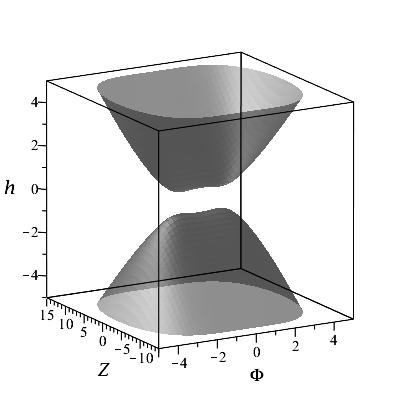}{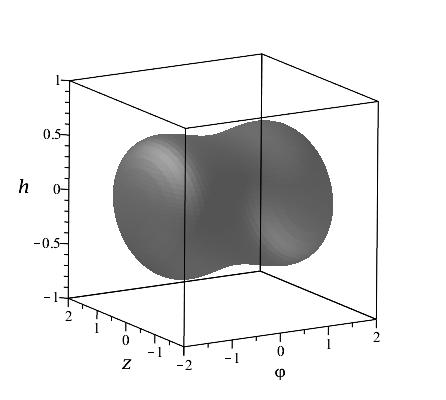}{\label{Sect0F3} Einstein - Higgs hypersurface projection in the classical subspace $\Sigma_\Phi$  -- two-sheeted hyperboloid with principal $Oh$ axis. Type $I_\Phi$.}{\label{Sect0p3} Einstein - Higgs hypersurface projection in the classical subspace $\Sigma_\Phi$  is a deformed ellipsoid. Type $III_\varphi$.}

\subsubsection*{Type $\mathbf{II}_\Phi$: simply connected Einstein-Higgs hypersurface: $\Sigma_\Phi$ $\mathbf{P}=[1,1,1,1,0.1,0.01]$;\\ $\mathbf{I}=[0.2,0.2,0.2,0.72059,\pm1]$:  $\alpha_mC_{\Phi}\approx 2\cdot10^{-6}\approx 0$)}

\TwoFigs{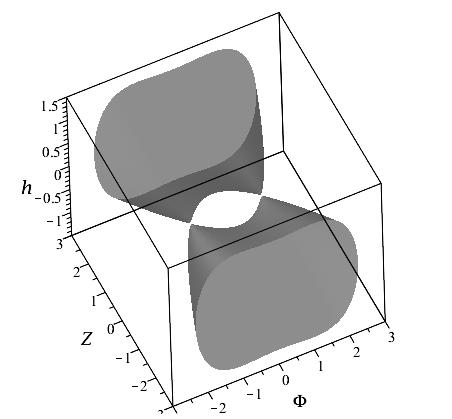}{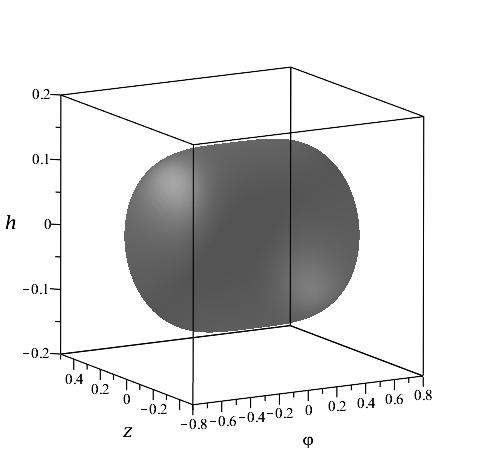}{\label{Sect3F3} Einstein - Higgs hypersurface projection in the classical subspace $\Sigma_\Phi$ is a one-sheet hyperboloid with principal $OZ$ axes.  In section 2 points $\mathbf{S}=\{A,B\}$. Type $II_\Phi$.}{\label{Sect3p3} Einstein - Higgs hypersurface projection in the classical subspace $\Sigma_\Phi$  is a deformed ellipsoid. Type $III_\varphi$.}
Note that in this case, in the vicinity of the plane $Z=0$ , the projection of the Einstein - Higgs hypersurface onto the classical subspace $\Sigma_\Phi$ degenerates into 2 pairs of cones with common vertices, which provide simply connectedness.

\subsubsection*{Type $\mathbf{III}_\Phi$: simply connected Einstein-Higgs hypersurface: $\Sigma_\Phi$ $\mathbf{P}=[1,1,1,1,0.1,0.01]$;\\ $\mathbf{I}=[0.2,0.2,0.2,0.2,\pm1]$:  $\alpha_mC_{\Phi}=.2396269975<1/4$)}

\ThreeFig{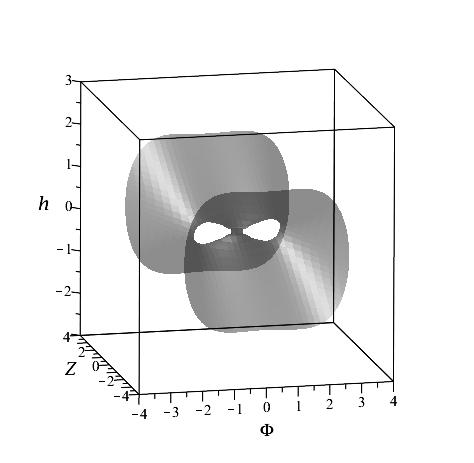}{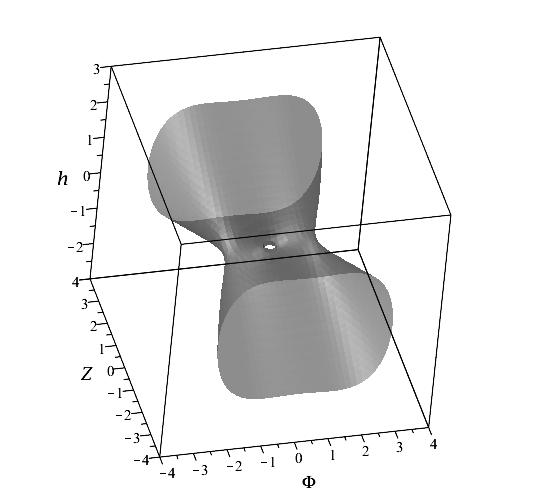}{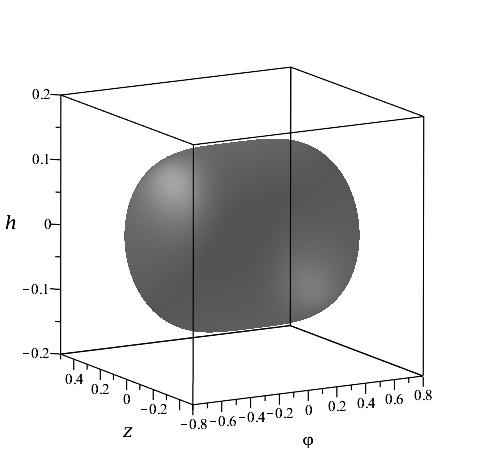}{\label{Sect2F3} Einstein - Higgs hypersurface projection in the classical subspace $\Sigma_\Phi$  is a one-sheet hyperboloid with the principal $OZ$ axis. $\mathbf{S}=\{\mathbf{E}_a,\mathbf{E}_b\}$ are two deformed ellipses. Type $III_\Phi$.}{\label{Sect2Fh3} Einstein - Higgs hypersurface projection in the classical subspace $\Sigma_\Phi$. From this perspective, an ellipsoidal gap is visible in the plane $h=0$.}{\label{Sect2p3} Einstein - Higgs hypersurface projection in the phantom subspace $\Sigma_\varphi$ is a deformed ellipsoid. $\mathbf{S}=\mathbf{E}$ is a deformed ellipse. Type $III_\varphi$.}
\subsubsection*{Type $\mathbf{IV}_\Phi$: simply connected Einstein-Higgs hypersurface: $\Sigma_\Phi$ \\
$\mathbf{P}=[1,1,1,1,0.1,0.02037300251]$; $\mathbf{I}=[0.2,0.2,0.2,0.2,\pm1]$:  $\alpha_mC_{\Phi}=1/4$)}
\TwoFigs{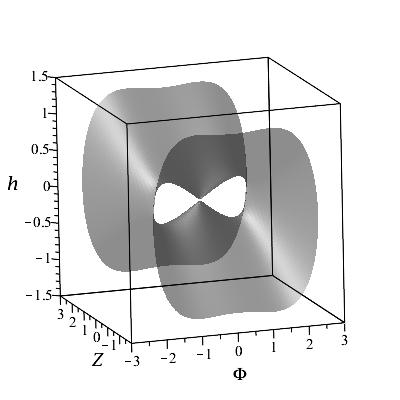}{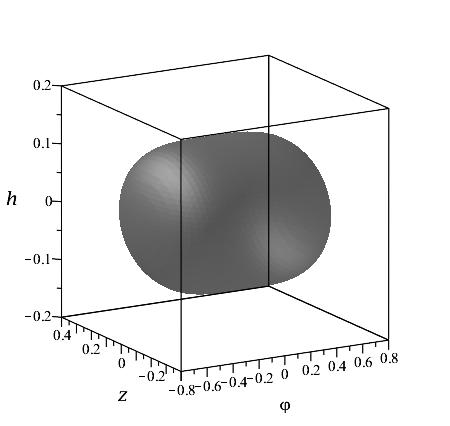}{\label{Sect4F3} Einstein - Higgs hypersurface projection in the classical subspace $\Sigma_\Phi$ is a one-sheet hyperboloid with principal $OZ$ axis. Figure eight in section $\mathbf{S}=\mbox{\large\bf $\infty$}$. Type $IV_\Phi$.}{\label{Sect4p3} Einstein - Higgs hypersurface projection in the classical subspace $\Sigma_\Phi$ is a deformed ellipsoid. Type $III_\varphi$.}
\subsubsection*{Type $\mathbf{V}_\Phi$: simply connected Einstein-Higgs hypersurface: $\Sigma_\Phi$ $\mathbf{P}=[1,1,1,1,1,0.1]$;\\ $\mathbf{I}=[0.2,0.2,0.2,0.2,\pm1]$:  $\alpha_mC_{\Phi}=0.3496>1/4$)}
\TwoFigs{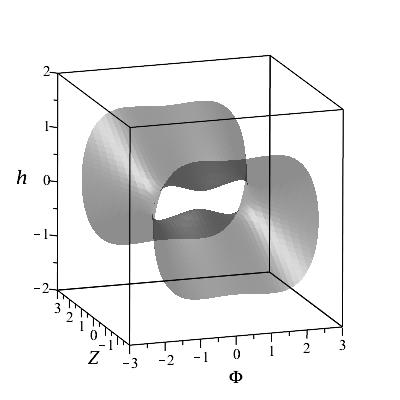}{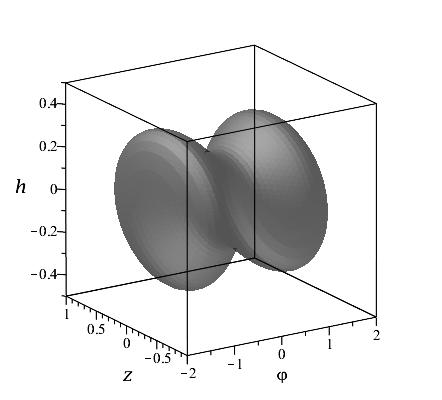}{\label{Sect1F3} Einstein - Higgs hypersurface projection in the classical subspace $\Sigma_\Phi$   is a one-sheet hyperboloid with the principal $OZ$ axis. Type $V_\Phi$.}{\label{Sect1p3} Einstein - Higgs hypersurface projection in the classical subspace $\Sigma_\Phi$ is a deformed ellipsoid. Type $III_\varphi$.}

\subsection{Conditions for the possibility of transitions of a dynamic system between the states of compression and expansion}
In order for the phase trajectory $ \mathbf{r} (\tau) $ to connect a pair of points $ M^- $ and $ M^+ $ with a negative and positive coordinate value $ h = h_- <0 $ and $ h = h_+> 0 $ respectively, it is necessary, first, that the regions of the Einstein-Higgs hypersurface with positive and negative values of $ h $ constitute a simply connected surface. Therefore, in models with $ \alpha <0$, $e = -1 $, which corresponds to the surface type $ \mathbf{I}_\Phi $ (Fig. \ref{Sect0F3} and \ref{Sect0p3}), such transitions are not possible.

When analyzing the behavior of phase trajectories, it is necessary to keep in mind Note 1, according to which, in the course of the evolution of the dynamical system, changes in the types of projections of the Einstein hypersurface are possible. For example, transitions of the kind $\mathbf{II}_\Phi \Leftrightarrow \mathbf{III}_\Phi$ are possible (Fig. \ref{Sect0F3}, \ref{Sect3F3}). Therefore, the transitions $ h_- \leftrightarrow h _ + $, which are impossible in the initial state of a dynamical system, may turn out to be possible during its evolution, and conversely.

\section{Phase trajectories and parameters of the dynamic system}
\hspace{12pt} Proceeding to the description and analysis of the phase trajectories of the dynamical system \eqref {Phi'} - \eqref{h'}, we will make some preliminary notes. \\[4pt]
{\note Since the system of equations \eqref{Phi'} - \eqref{h'} is autonomous, it is invariant with respect to arbitrary time shifts
\begin{equation}\label{t->t}
\tau\to \tau+\mathrm{Const}, \quad (\forall\ \mathrm{Const}).
\end{equation}
This makes it possible to choose any moment of time as the starting moment of time. We will everywhere assume $ \tau_0 = 0 $ as the initial moment of time. At that, we can investigate the evolution of a dynamical system for negative values of the variable $ \tau $.}\\[4pt]
{\note As we noted above, the initial conditions for the dynamical system (\ref{Phi'}) - (\ref{z'}) must satisfy the integral condition (\ref{EqEinst0_4_E}), whence we get the initial value of the function $ h(0) $ \eqref{h0}.}\\[4pt]
{\note If for $ \lambda> 0 $ there are attractive point $M^+_{0,0} $ in the subspace $ \Sigma_ \Phi $ for $ h> 0 $ and repulsive point $M^-_{0,0} $ for $ h <0 $, then phase trajectories with initial conditions corresponding to $ h> h_- $ or $ h <h_- $ exhibit the standard behavior: $ h (\infty) \to h_+ $ or $ h(\infty ) \to h_- $, respectively, i.e. tend to expand at a constant speed or contract at a constant speed without changing the phase. The same phase trajectories, the initial value of the Hubble constant of which is within the open interval $ h\in (h_-, h_+) $, can pass through the position $ h = 0 $ and thus make transitions between the stages of compression and expansion. In the case of $ \lambda <0 $ and the absence of these singular points, the same role can perform the singular points $M^\pm_{\pm1,\pm1}, M^\pm_{\pm1,0}, M^\pm_{0,\pm1}$.}\\[4pt]
{\note We will represent three-dimensional projections of the phase trajectories of the dynamical system in the subspaces $ \Sigma_\Phi $ and $ \Sigma_\varphi $ together with the projections of the Einstein - Higgs hypersurface in these subspaces. In this case, we should remember Note 1, according to which the projections of this hypersurface change with time, therefore, in the case of a complete doublet, the projection of the phase trajectory does not have to lie on the projection of this hypersurface except for one point.}\\[4pt]
{\note We will group the results according to the types of global behavior of the dynamical system, keeping in mind the main goal of our study - to study the possibilities of changing the phases of cosmological expansion and contraction.}

\subsection{Examples of singlets with changing phases of expansion and contraction}
To understand the trend in the evolution of the physical parameters of the cosmological model, let us first consider, for simplicity, the evolution of scalar singlets. Recall that in cases that reduce to a singlet, the projection of the Einstein - Higgs hypersurface does not change with time; therefore, in these cases, the projections of the phase trajectories lie on the corresponding projections of the Einstein - Higgs hypersurface. \\

\textbf{Classic singlet: $\mathbf{P}=[1,1,1,1,1,0.1]$; $\mathbf{I}=[1,0.1,0,0,1]$};  principal  $OZ$ axis -- the first type of endless rolling and $\mathbf{P}=[-1,1,-1,1,1,0.1]$; $\mathbf{I}=[0.3,0,0,0,1]$; principal  $Oh$ axis -- second type of endless rolling. \\
\TwoFigsReg{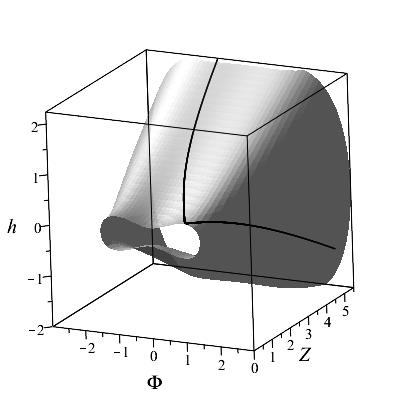}{6}{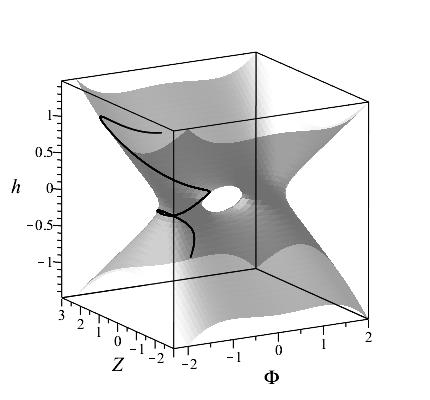}{6}{\label{Fase1_c_pm} The first type of phase trajectory of the dynamical system in the subspace $\Sigma_\Phi$.}{\label{Fase2_c_pm} The second type of phase trajectory of the dynamical system in the subspace $\Sigma_\Phi$.}
In Fig. \ref{Fase1_c_pm} - \ref{E12_c_pm} shows two types of the process of changing the phase of cosmological expansion to contraction for a classical singlet.
Thus, in the case of a scalar singlet, firstly, the transitions $h_+ \to h_- $ \cite {Yu_20_1} are possible, and secondly, in the transition domain $ h = 0 $ the effective energy tends to zero $ \mathcal{E}_m \to 0 $, which was noticed in early works (see, for example, \cite{Yu_19_2}). Note also that according to \eqref{omega} in the domain $h\to 0$ for the classical singlet $\Omega\to-\infty$.
\TwoFigs{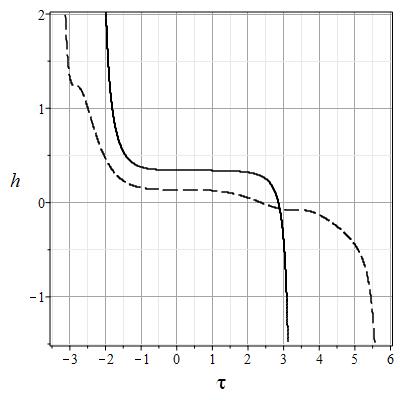}{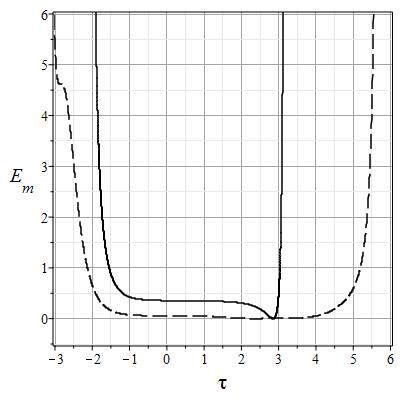}{\label{h12_c_pm} Evolution of the Hubble constant $h(\tau)$. Solid line -- first type, dotted line -- second type.}{\label{E12_c_pm} Evolution of the Hubble constant $\mathcal{E}_m$. Solid line -- first type, dotted line -- second type.}
\textbf{Phantom singlet: $\mathbf{P}=[1,1,1,1,1,0.1]$; $\mathbf{I}=[0,0,0.001,0.001,-1]$}\\[6pt]
In Fig. \ref{h1_p_mp} -- \ref{Fase1_p_mp} shows the process of changing the phase of cosmological compression to expansion for a phantom singlet.\\[6pt]
\TwoFigs{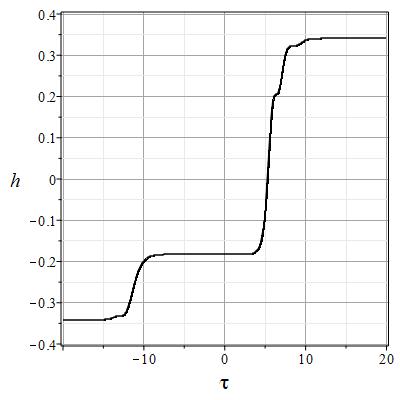}{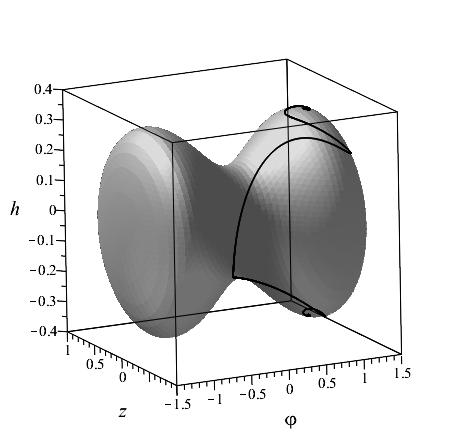}{\label{h1_p_mp}Evolution of the Hubble constant $h(\tau)$. }{\label{Fase1_p_mp}The phase trajectory of the dynamical system in the subspace $\Sigma_\varphi$.}

Finally, in Fig. \ref{O12_F_pm} -- \ref{O1_p_mp} the evolution of the cosmological acceleration is shown for the scalar singlets considered above.\\[6pt]
{\note Note that the transition of the dynamical system through the position $ h = 0 $, firstly, corresponds to a change in the expansion and contraction phases, and, secondly, to infinite values of the invariant cosmological acceleration in accordance with the formula \eqref{omega}. In this case, however, the scale factor $ a(\tau) $ and its first two derivatives $ \dot{a} $ and $ \ddot{a} $ remain continuous and bounded functions, at least of the class $\mathrm{C}^2$. Finally, the invariant curvature \eqref{sigma} tends to a finite value as $ h\to 0 $.}
\begin{equation}\label{sigma->}
\lim\limits_{h\to0}\sigma=\sigma_0=\sqrt{6\dot{H}^2}<\infty.
\end{equation}
\TwoFigsReg{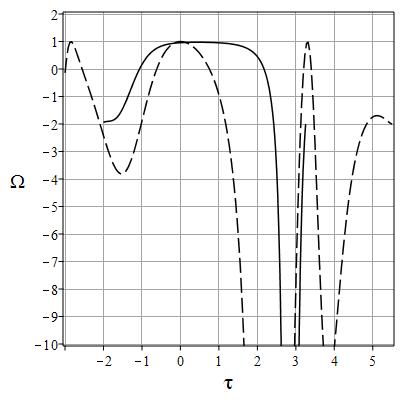}{6}{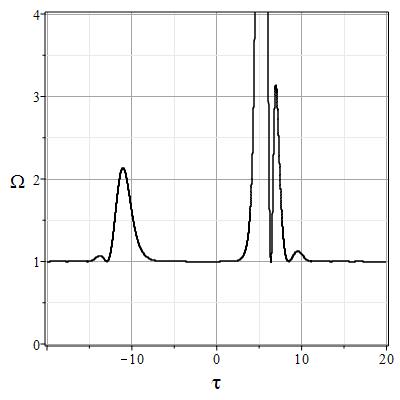}{6}{\label{O12_F_pm} Evolution of invariant cosmological acceleration  $\Omega(\tau)$ for classic singlet. Solid line -- first type, dotted line -- second type.}{\label{O1_p_mp} Evolution of invariant cosmological acceleration $\Omega(\tau)$ for phantom singlet.}

It can be seen from these figures that if the classical singlet only at times reaches the state of contraction or expansion with constant velocity $ \Omega = 1, H = \mathrm{Const} $, then for the phantom singlet this state is dominant.\\[6pt]

\textbf{Doublet with a constant classical field: $\mathbf{P}=[1,1,1,1,1,1],\; \mathbf{I}=[1,0,0.1,0,1]$}\\[6pt]
We place the coordinates of the classical scalar field at the attracting point $\Phi=\Phi_+=1$.
In Fig. \ref{hF=1} -- \ref{Fase_F=1} shows the change in the phase of cosmological compression to expansion for this doublet, and in Fig. \ref{O_F=1} -- \ref{E_F=1} shows the evolution of the cosmological acceleration and effective energy for the considered scalar doublet with a constant classical field.\\[6pt]
\TwoFigs{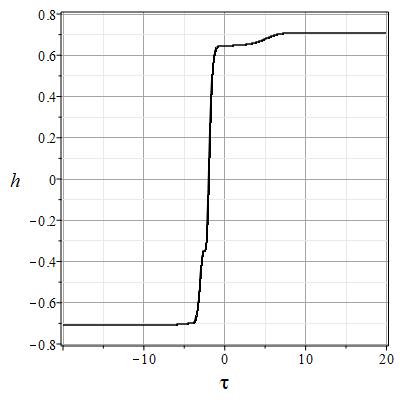}{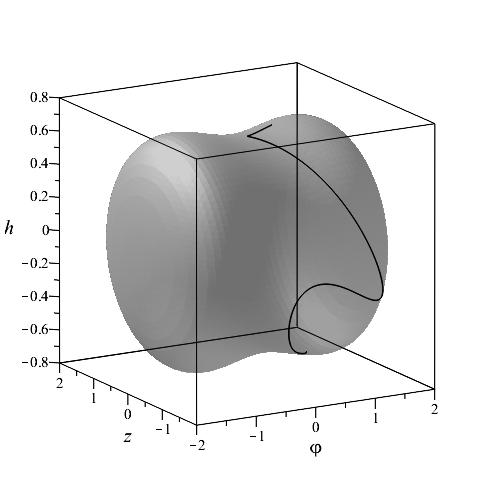}{\label{hF=1}Evolution of the Hubble constant $h(\tau)$. }{\label{Fase_F=1} Phase trajectory of the dynamical system in the subspace $\Sigma_\varphi$ for $\Phi=\Phi_+=1$.}
\TwoFigs{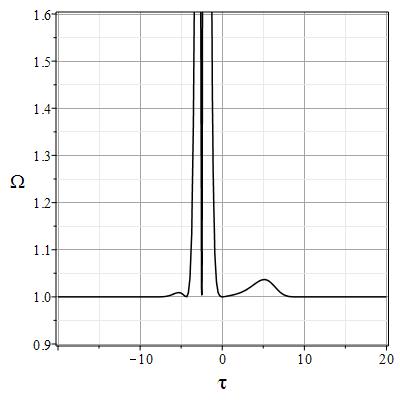}{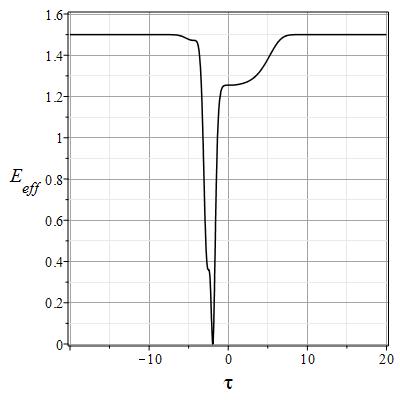}{\label{O_F=1}Evolution of invariant cosmological acceleration $\Omega(\tau)$ for the scalar doublet with a constant classical field.}{\label{E_F=1}The evolution of the effective energy  $\mathcal{E}_m(\tau)$ for the scalar doublet with a constant classical field.}

\textbf{Doublet with a constant phantom field: $\mathbf{P}=[1,1,1,1,1,0.1],\; \mathbf{I}=[1.5,0,1,0,1]$}\\[6pt]
Place the coordinates of the phantom scalar field in the attracting point $\varphi=\varphi_+=1$.
In Fig. \ref{hp=1} - \ref{Fase_p=1} shows the process of changing the phase of cosmological compression to expansion for this doublet, and in Fig. \ref{O_p=1} -- \ref{E_p=1} shows the evolution of cosmological acceleration and effective energy for the considered scalar doublet with a constant phantom field.

Comparison of pairs of similar plots $ h (\tau) $, $ \Omega (\tau) $, etc. for scalar singlets and scalar doublets with one constant field:
Fig. \ref{h12_c_pm} $\leftrightarrow$ \ref{hp=1}, \ref{hF=1} $\leftrightarrow$ \ref{h1_p_mp}, \ref{O12_F_pm} $\leftrightarrow$ \ref{O_p=1},
\ref{O1_p_mp} $\leftrightarrow$ \ref{O_F=1}, etc., reveals the related nature of their behavior, which confirms and supplements with details Statement 5.\\
\TwoFigs{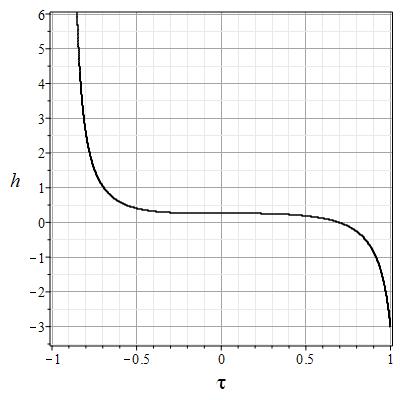}{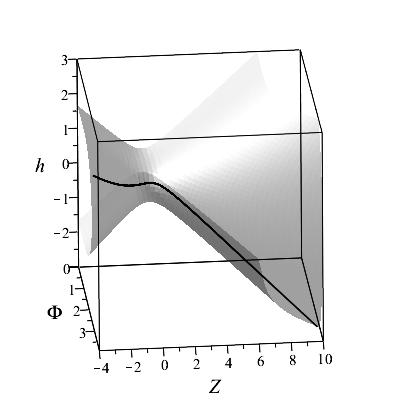}{\label{hp=1}Evolution of the Hubble constant $h(\tau)$. }{\label{Fase_p=1}Phase trajectory of the dynamical system in the subspace $\Sigma_\Phi$ for $\varphi=\varphi_+=1$.}
\TwoFigs{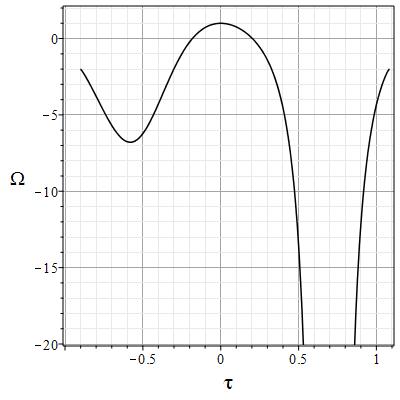}{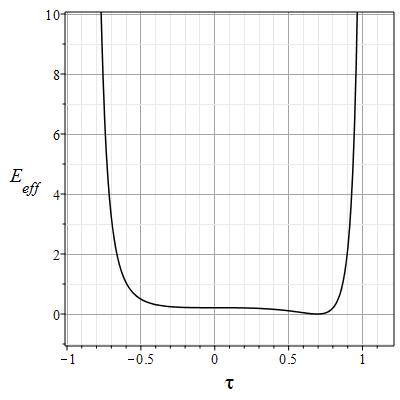}{\label{O_p=1}Evolution of invariant cosmological acceleration  $\Omega(\tau)$ for scalar doublet with constant phantom field.}{\label{E_p=1}The evolution of the effective energy   $\mathcal{E}_m(\tau)\equiv \mathcal{E}_{eff}$ for scalar doublet with constant phantom field.}
\subsection{Examples of a scalar doublet with a change in the expansion and contraction phases}

\textbf{Full doublet with multiple transitions $\mathbf{P}=[1,1,1,1,1,1],\; \mathbf{I}=[0.2,0,0.2,0,1]$}\\[6pt]
In Fig. \ref{h_fp_1} - \ref{Fase_fp_P1} shows the process of changing the phase of cosmological contraction to expansion and back for this doublet, and in Fig. \ref{O_fp_1} -- \ref{E_fp_1} shows the evolution of the cosmological acceleration and effective energy for the considered scalar doublet with a constant classical field.

In the graphs above, we observe a chain of 4 transitions $h_+\to h_-\to h_+\to h_-\to h_+$ (expansion - contraction - expansion - contraction - expansion). The triple transition of the dynamical system through the point $ h = 0 $ corresponds to infinite bursts of cosmological acceleration and zero (or close to zero) values of the effective energy $\mathcal{E}_{eff}\equiv \mathcal{E}_m$. After the end of the oscillations, the value of the effective energy tends to a constant value, and the invariant cosmological acceleration $ \Omega $ tends to a value corresponding to the expansion: with a constant velocity $\Omega\to1, H\to H_0>0$.

\begin{flushleft}
\begin{tabular}{lcr}
\hspace{-1cm}
\parbox{5.5cm}{\includegraphics[width=5.5cm]{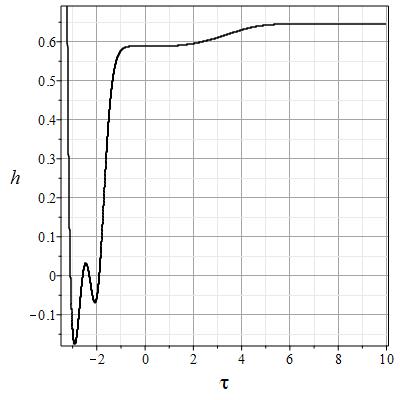}}  & \hspace{-0.5cm} \parbox{5.5cm}{\includegraphics[width=5.5cm]{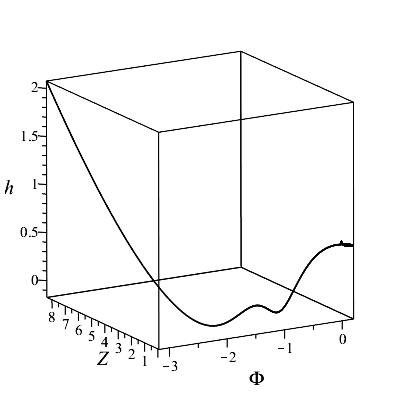}}  & \hspace{-1.5cm}\parbox{5.5cm}{\includegraphics[width=5.5cm]{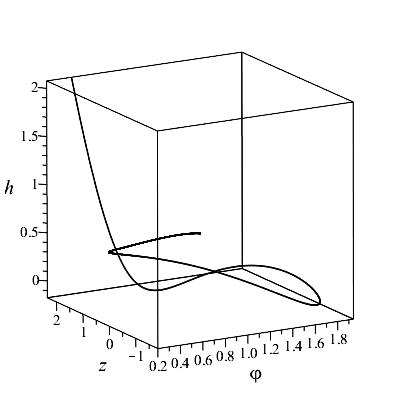}}
\\
\hspace{-1cm}
\parbox{5cm}{\vspace{7pt}\refstepcounter{figure}Fig. \thefigure.\quad\label{h_fp_1}Evolution of the Hubble constant $h(\tau)$. \vfill} &\hskip -10pt \parbox{5cm}{\vspace{7pt}\refstepcounter{figure} Fig. \thefigure.\quad \label{Fase_fp_F1}The phase trajectory of the dynamical system in the subspace $\Sigma_\Phi$.\vfill} & \parbox{5cm}{\vspace{7pt}
\refstepcounter{figure}Fig. \thefigure.\quad \label{Fase_fp_P1}The phase trajectory of the dynamical system in the subspace $\Sigma_\varphi$.\vfill}\\
\end{tabular}
\end{flushleft}
\vspace{7pt}

Note that the solution in this case, at times less than $ \tau = -3.4 $, rests on the point with the coordinate $ h \approx 7 $, which is associated with the above-mentioned global change in the topology of the Einstein-Higgs hypersurface in the subspace $ \Sigma_ \Phi $ -- violation of its simple connectivity at this moment in time (Fig. \ref{Fig_Rec1}).  In Fig. \ref{Fig_Rec1} shows the phase trajectory in the subspace $\Sigma_\Phi$ and the projection of the Einstein-Higgs hypersurface onto this subspace at the time instant  $\tau=-3.4$. The starting point of the trajectory lies on the vertex of the left side of the two-sheet hyperboloid. Over time, the simply connected of the Einstein - Higgs hypersurface is restored, and the model asymptotically enters the expansion mode with constant velocity $\Phi\to \mathrm{Const},\ \varphi\to \mathrm{Const}$, see Fig. \ref{Fig_Rec2} -- \ref{Fig_Rec3}.

\begin{flushleft}
\begin{tabular}{lcr}
\hspace{-1cm}
\parbox{5.5cm}{\includegraphics[width=5.5cm]{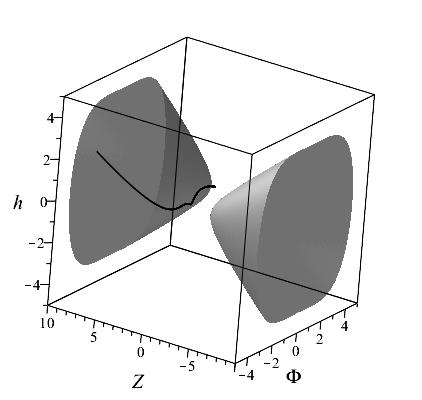}}  & \hspace{-0.5cm} \parbox{5.5cm}{\includegraphics[width=5.5cm]{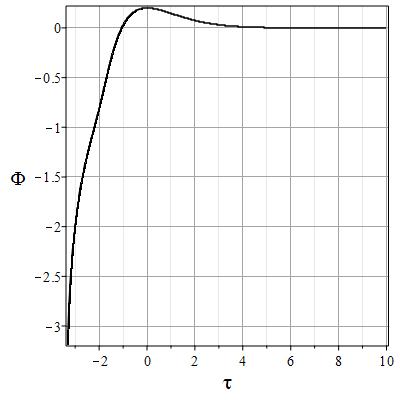}}  & \hspace{-1.5cm}\parbox{5.5cm}{\includegraphics[width=5.5cm]{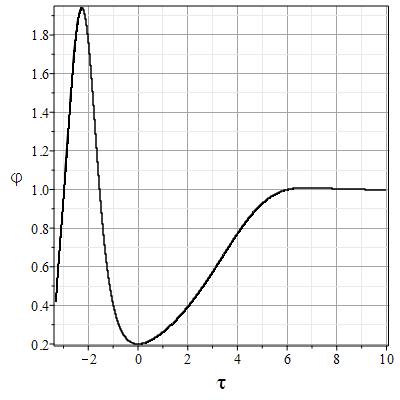}}
\\
\hspace{-1cm}
\parbox{5cm}{\vspace{7pt}\refstepcounter{figure}Fig. \thefigure.\quad\label{Fig_Rec1} The phase trajectory of a dynamical system in the subspace $\Sigma_\Phi$, superimposed on the projection of the Einstein - Higgs hypersurface at time $\tau=-3.4$.   \vfill} &\hskip -10pt \parbox{5cm}{\vspace{7pt}\refstepcounter{figure}Fig. \thefigure.\quad \label{Fig_Rec2}Evolution of the potential of the classical field $\Phi(\tau)$.\vfill} & \parbox{5cm}{\vspace{7pt}
\refstepcounter{figure} Fig. \thefigure.\quad \label{Fig_Rec3}Evolution of the potential of the phantom field  $\varphi(\tau)$.\vfill}\\
\end{tabular}
\end{flushleft}
\vspace{7pt}

\TwoFigs{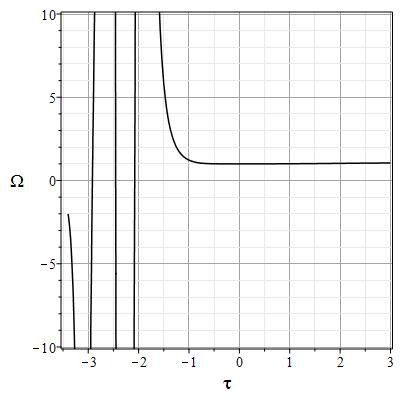}{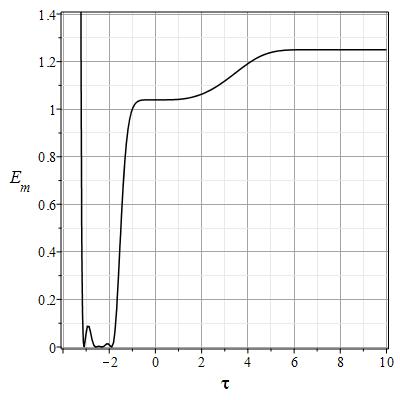}{\label{O_fp_1}Evolution of invariant cosmological acceleration  $\Omega(\tau)$ for the full scalar doublet.}{\label{E_fp_1}The evolution of the effective energy   $\mathcal{E}_m(\tau)$ for the full scalar doublet.}

\subsection{Examples transitions inflation compressed mode on inflation extension mode}
In the case when there are attractive singular points with coordinates $ h_- <0 $ and $ h_+>0 $, transitions between these points become possible from the inflationary compression regime to the expansion regime with a constant velocity ($ \Omega=1 $), and backwards. These modes of evolution are possible under conditions of dominance of the phantom field or the classical one, respectively. Due to the importance of this type of behavior, we will consider it in detail. \\[6pt]
\textbf{Full doublet: $\mathbf{P}=[1,1,1,1,0.5,-0.01],\; \mathbf{I}=[0.2,0.1,0.1,0.1,1]$}\\[6pt]
Let's write down the coordinates of the singular points (the signs take on independent values):
\begin{eqnarray}
M^\pm_{\pm1,0}=[\pm1,0,0,0,\pm 0.283];\hskip 20pt & M^\pm_{0,\pm1}=[0,0,\pm0.5,0,,\pm 0.434]; \nonumber\\
M^\pm_{\pm1,\pm1}=[\pm1,0,\pm0.5,0,\pm0.292]. & \nonumber
\end{eqnarray}
Thus, in our case, there are 16 singular points, the points $M^\pm_{0,0}$ are absent. Thus, 4 points of $M^\pm_{\pm1,0}$ in the subspace $\Sigma_\Phi$ and in the subspace $\Sigma_\varphi$ re saddle points, 4 points $M^\pm_{0,\pm1}$in the half-space $h>0$ are attracting in both subspaces, in the half-space $h<0$ are repulsive, and finally 8 points $M^\pm_{\pm1,\pm1}$ in the half-space $h>0$ are attracting in both subspaces and in the half-space $h<0$ are repulsive in both subspaces. The abundance of points and their characteristics is due to enough complex behavior of phase trajectories (Fig. \ref{hFZ_up} and \ref{hpz_up}).
\TwoFigs{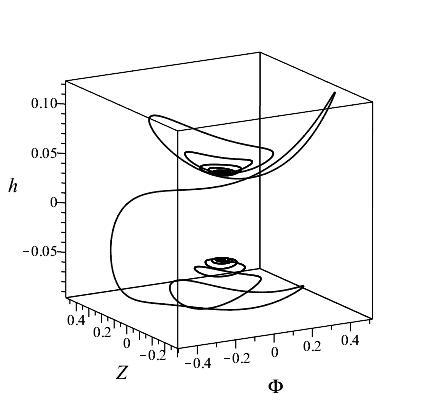}{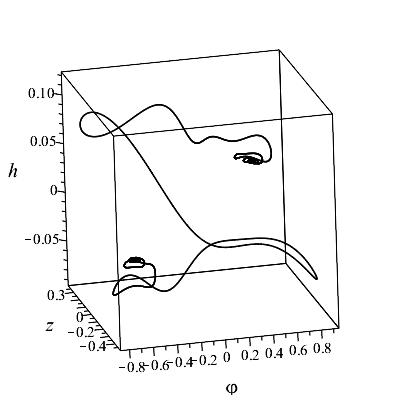}{\label{hFZ_up} Phase trajectory of the classic field in the subspace $\Sigma_\Phi$ for parameters $\mathbf{P}=[1,1,1,1,0.5,-0.01]$ and initial conditions $\mathbf{r}_0=[0.2,0.1,0.1,0.1,1]$.}{\label{hpz_up}Phase trajectory of the phantom field in the subspace $\Sigma_\varphi$ for parameters $\mathbf{P}=[1,1,1,1,0.5,-0.01]$ and initial conditions $\mathbf{r}_0=[0.2,0.1,0.1,0.1,1]$.}

The phase trajectories leave the neighborhood of repulsive singular points in the half-space $ h<0 $ and, as a result of a rather complicated transition to the half-space $ h>0 $, tend to the attracting singular points. In this case, the phase trajectories go around from different sides many repulsive and saddle singular points that occur on their way and are indicated above. In Fig. \ref{h_up} -- \ref{O_Fp_up} shows the evolution of physical parameters for this doublet.

\ThreeFigReg{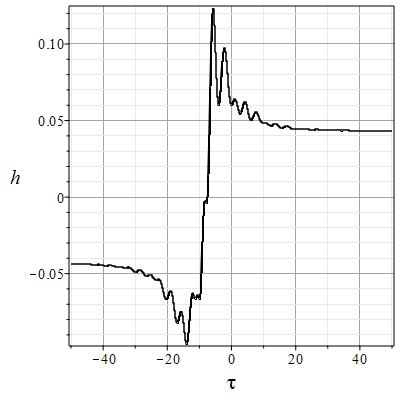}{5}{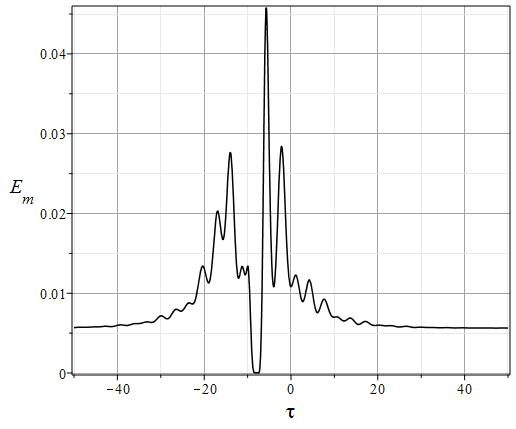}{5}{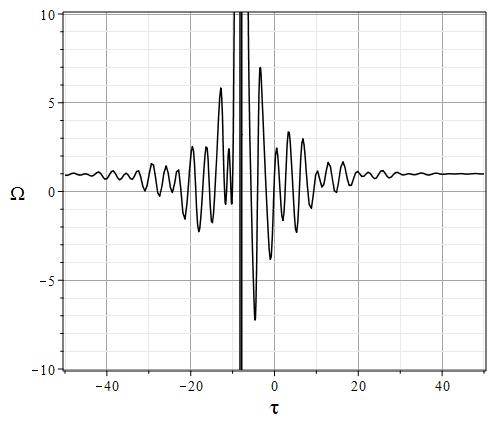}{5}{\label{h_up}Evolution of the Hubble constant $h(\tau)$. }{\label{E_Fp_up}The evolution of the effective energy $\mathcal{E}_m$}{\label{O_Fp_up} Evolution of  cosmological acceleration $\Omega$}
\textbf{Full doublet with $\lambda>0$ and phantom dominance: $\mathbf{P}=[1,1,1,1,0.5,0.02],\; \mathbf{I}=[0.01,0.01,0.2,0.1,1]$}\\[6pt]
Let's write down the coordinates of the singular points (the signs take on independent values):
\begin{eqnarray}
M^\pm_{0,0}=[0,0,0,0,\pm0.0817]; &  M^\pm_{\pm1,0}=[\pm1,0,0,0,\pm 0.309]; & \nonumber\\
M^\pm_{0,\pm1}=[0,0,\pm0.5,0,\pm 0.309]; & M^\pm_{\pm1,\pm1}=[\pm1,0,\pm0.5,0,\pm0.309]. & \nonumber
\end{eqnarray}

Thus, in our case, there are 18 singular points. The singular points $M^\pm_{0,0}$ in the subspace $\Sigma_\Phi$ for $h>0$ are attracting, and for $h<0$ re repulsive, in the subspace $\Sigma_\varphi$ are saddle points, 4 points $M^\pm_{\pm1,0}$ are saddle points everywhere, 4 points $M^\pm_{0,\pm1}$ in the half-space $h>0$ re attracting in both subspaces, in the half-space $h<0$ are repulsive, finally, 8 points $M^\pm_{\pm1,\pm1}$ in the subspace $\Sigma_\Phi$ are saddle points, and the subspace $\Sigma_\varphi$ in the half-space $h>0$ are attracting and in the half-space $h<0$ are repulsive. The phase trajectories of the dynamic system are shown in Fig. \ref{Fase3_c_mp} and \ref{Fase3_p_mp}.

The phase trajectories leave the neighborhood of the repulsive singular points $M^-_{0,0}$ in the half-space $h<0$ and after passing to the half-space $h>0$ tend to the attracting singular points $M^+_{0,0}$. In this case, the cosmological model cross over from a constant velocity compression mode ($ H = H_0<0, \Omega = 1 $) to a constant velocity expansion mode ($ H = H_0>0, \Omega = 1 $). At the moment of transition, the effective energy is zero, and the invariant acceleration $\Omega\to-\infty$.

\TwoFigs{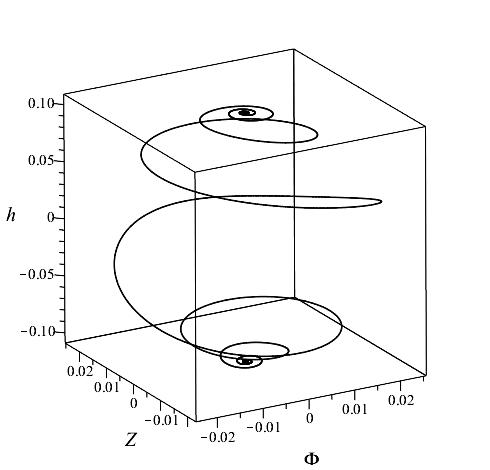}{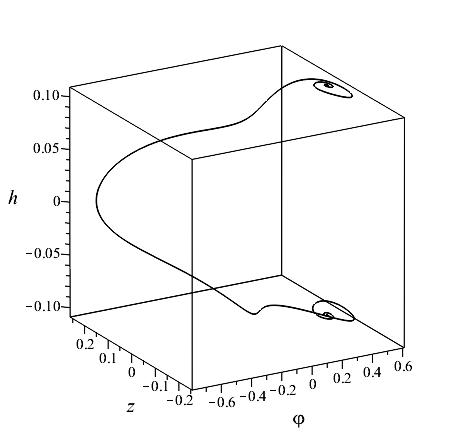}{\label{Fase3_c_mp} Phase trajectory of the classic field in the subspace $\Sigma_\Phi$ for parameters $\mathbf{P}=[1,1,1,1,0.5,0.02]$ and initial conditions $\mathbf{I}=[0.01,0.01,0.2,0.1,1]$.}{\label{Fase3_p_mp} Phase trajectory of the phantom field in the subspace $\Sigma_\varphi$ for parameters $\mathbf{P}=[1,1,1,1,0.5,0.02]$ and initial conditions $\mathbf{I}=[0.01,0.01,0.2,0.1,1]$.}
In Fig. \ref{h_fp_3} -- \ref{O_fp_3} shows evolution of the Hubble constant,  cosmological acceleration and effective energy for the considered scalar doublet.\\[6pt]

\ThreeFigReg{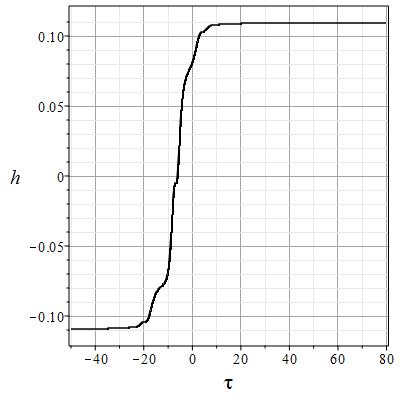}{5}{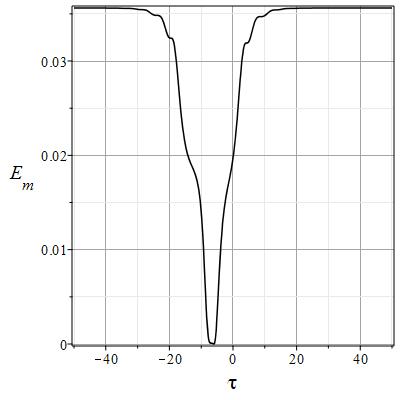}{5}{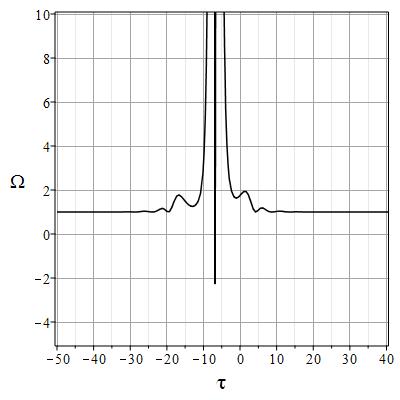}{5}{\label{h_fp_3}Evolution of the Hubble constant $h(\tau)$. }{\label{Figs/E_fp_3}The evolution of the effective energy $\mathcal{E}_m$}{\label{O_fp_3}Evolution of  cosmological acceleration $\Omega$}

\subsection{Example of oscillatory mode}
Oscillatory modes become possible in conditions of equal competition between the classical and phantom fields.\\[6pt]
\textbf{Full douplet with $\lambda<0$: $\mathbf{P}=[1,1,1,1,0.5,-0.02],\; \mathbf{I}=[0.2,0.1,0.1,0.1,1]$}\\[6pt]
Let's write down the coordinates of the singular points (the signs take on independent values):
\begin{eqnarray}
M^\pm_{\pm1,0}=[\pm1,0,0,0,\pm 0.286];\hskip 20pt & M^\pm_{\pm1,\pm1}=[\pm1,0\pm0.5,0,,\pm 0.286]; \nonumber
\end{eqnarray}
Thus, in our case, there are 12 singular points, the points $ M^pm_{0,0} $ and $ M^\pm_{0, \pm1} $ are absent. All singular points in the subspace $ \Sigma_\Phi$ are saddle points, the singular points $ M^\pm_{\pm1,\pm1}$ for $h>0$ are attracting, and for $h<0$ they are repulsive, the singular the points $ M^\pm_{\pm1,0}$ in the subspace $\Sigma_\varphi$ are saddle points. The phase trajectories of the dynamic system are shown in Fig. \ref{Figs/Fase_fp_4F} and \ref{Fase_fp_4P}. The projections of the Einstein-Higgs hypersurface are shown at a finite time.

\TwoFigs{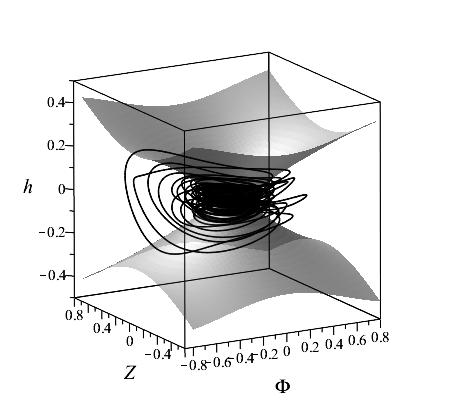}{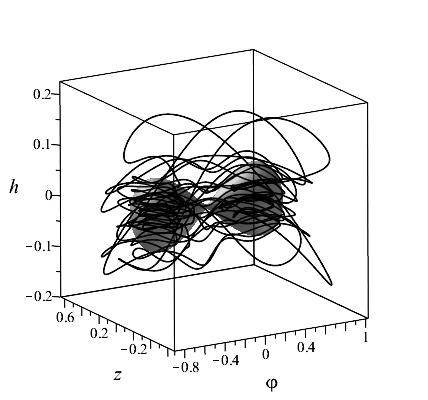}{\label{Figs/Fase_fp_4F} Phase trajectory of the classic field in the subspaces $\Sigma_\Phi$ for parameters $\mathbf{P}=[1,1,1,1,0.5,-0.02]$ and initial conditions $\mathbf{I}=[0.2,0.1,0.1,0.1,1]$.}{\label{Fase_fp_4P} Phase trajectory of the phantom field in the subspacesе $\Sigma_\varphi$ for parameters $\mathbf{P}=[1,1,1,1,0.5,-0.02]$ and initial conditions $\mathbf{I}=[0.2,0.1,0.1,0.1,1]$.}

In Fig. \ref{h_fp_4} -- \ref{O_fp_4} shows evolution of the Hubble constant,  cosmological acceleration and effective energy for the considered scalar doublet.\\[6pt]

\ThreeFigReg{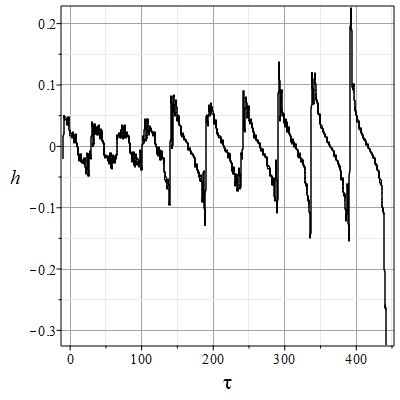}{5}{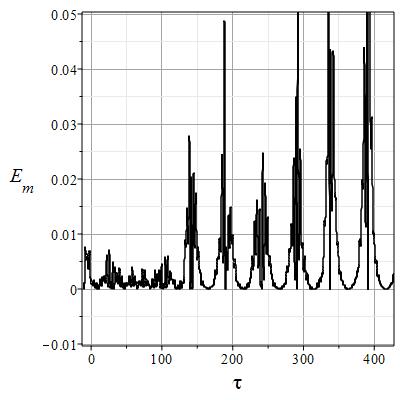}{5}{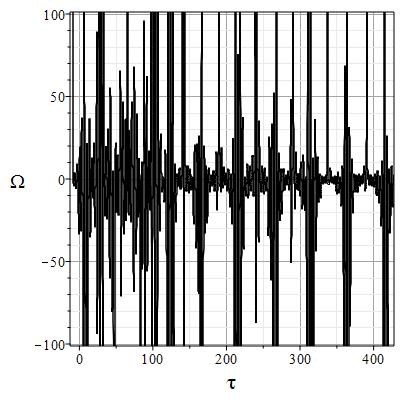}{5}{\label{h_fp_4}Evolution of the Hubble constant $h(\tau)$. }{\label{Figs/E_fp_4}The evolution of the effective energy $\mathcal{E}_m$}{\label{O_fp_4}Evolution of  cosmological acceleration $\Omega$}

\subsection{Example of modes with preservation and violation of the sign symmetry of the phantom field}
When constrictions occur and following gaps of the hypersurface in the subspace $\Sigma_\varphi$ similar to those shown in Fig. \ref{evol_fant_surf}, the phase trajectory of the phantom field may be trapped in one of the symmetrical parts. In this case, the potential of the phantom field can save its sign, or change it to the opposite. \\[6pt]
\textbf{Full doublet with $\lambda>0$: $\mathbf{P}=[0.1,0.1,1,1,1,0.01]$, $\mathbf{I_1}=[0.1,0.1,-0.1,0.1,1]$ and $\mathbf{I_2}=[0.1,0.1,-0.1,-0.1,1]$}\\[6pt]
In Fig. \ref{Fase_Fantom_asym} and \ref{Fase_Fantom_sym} show the phase trajectories of the dynamical system in the phantom subspace $\Sigma_\varphi$, and the projections of the hypersurfaces are shown at a finite time.

\TwoFigs{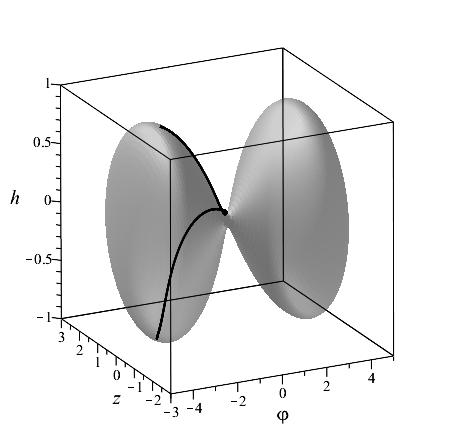}{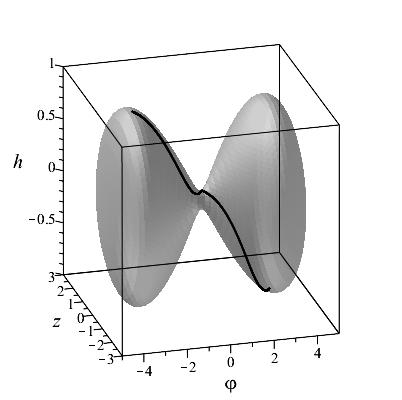}{\label{Fase_Fantom_asym} Phase trajectory of the phantom field in the subspace $\Sigma_\varphi$ for initial conditions $\mathbf{I_1}$.}{\label{Fase_Fantom_sym} Phase trajectory of the phantom field in the subspace $\Sigma_\varphi$ for initial conditions $\mathbf{I_2}$.}
Notice, that the parameters of the models presented on \ref{Fase_Fantom_asym} and \ref{Fase_Fantom_sym} completely coincide, and the initial conditions differ only in the sign of $ z_0 $.

In Fig. \ref{h_Fantom_asym} -- \ref{O_Fantom_asym} shows evolution of the Hubble constant, effective energy and cosmological acceleration for the considered scalar doublet with initial conditions $\mathbf{I_1}$, and in Fig. \ref{h_Fantom_sym} -- \ref{O_Fantom_sym} shows similar plots in case of initial conditions $\mathbf{I_2}$.

We see that the evolution of all physical characteristics of these two systems exhibits the same behavior with a certain time delay in the first case, although the sign of the phantom field potential is different.

\ThreeFigReg{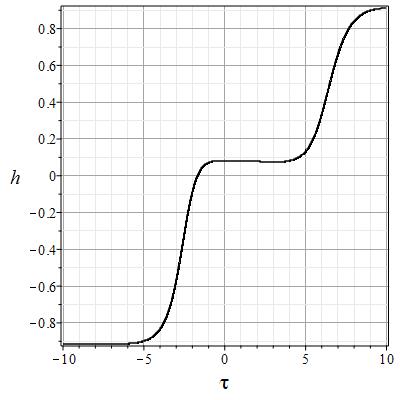}{5}{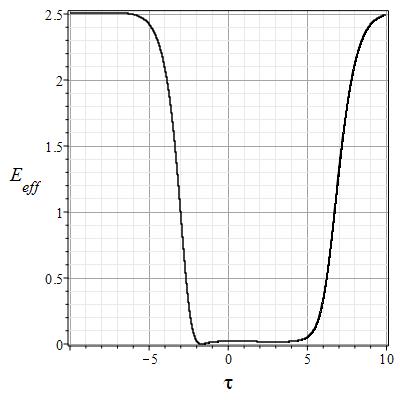}{5}{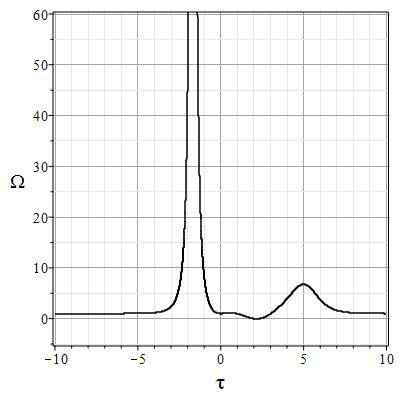}{5}{\label{h_Fantom_asym}Evolution of the Hubble constant $h(\tau)$ for initial conditions $\mathbf{I_1}$.}{\label{E_Fantom_asym}The evolution of the effective energy $\mathcal{E}_m$ for initial conditions $\mathbf{I_1}$}{\label{O_Fantom_asym}Evolution of cosmological acceleration $\Omega$ for initial conditions $\mathbf{I_1}$}
\ThreeFigReg{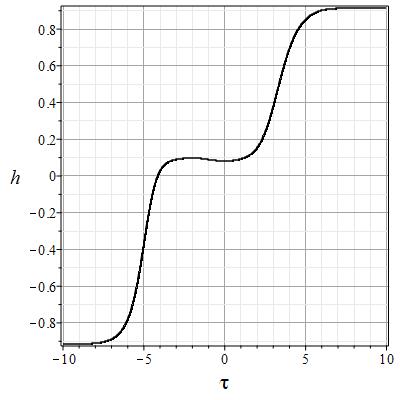}{5}{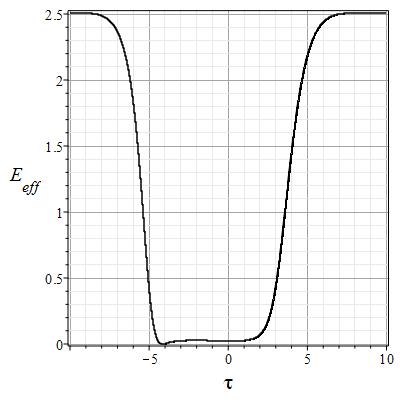}{5}{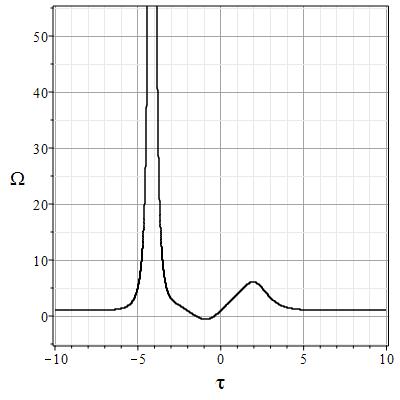}{5}{\label{h_Fantom_sym}Evolution of the Hubble constant $h(\tau)$ for initial conditions $\mathbf{I_2}$.}{\label{E_Fantom_sym}The evolution of the effective energy $\mathcal{E}_m$ for initial conditions $\mathbf{I_2}$}{\label{O_Fantom_sym}Evolution of cosmological acceleration $\Omega$ for initial conditions $\mathbf{I_2}$}
\section{Conclusion}
So, let us note the closeness of the methods for studying quintom models in the work \cite{Leon18} and in our work in relation to the use of the apparatus of the qualitative theory of differential equations.  In addition to the methods of the qualitative theory of differential equations, in this work, the study of the internal symmetry of solutions and their singular analysis were used as additional methods.  In our work, such an additional method is the study of the topology of the Einstein - Higgs hypersurface, the simply connectedness of three-dimensional projections of which is violated in a number of cases, exerting a significant influence on the behavior of the dynamical system.  Comparison of the results of our work with the results of the work \cite{Leon18} is not possible, since the forms of the potentials of the quintom in these works are significantly different: in the work \cite{Leon18} is an exponential potential containing even and odd orders of scalar fields, in our work –– this is the Higgs potential, containing only even orders of fields: 0.2 and 4.

Indeed, performing the \emph{bijective} transformation $\{\phi,\psi\} \leftrightarrow \{\Phi,\Psi\}$ on the work potential \cite{Leon18} of the form
\[ \Phi=\mathrm{e}^{-\sqrt{6}m\phi}\geqslant 0,\; \Psi=\mathrm{e}^{-\sqrt{6}n\psi}\geqslant 0\Leftrightarrow
\phi=-\frac{1}{m\sqrt{6}}\ln\Phi,\; \psi=-\frac{1}{n\sqrt{6}}\ln\Psi,
\]
we bring the corresponding equation of the Einstein hypersurface to the form of an algebraic equation of the second order:
\[ 3H^2-\frac{Z^2}{2}+\frac{z^2}{2}-V_0\Phi\Psi+\Lambda=0,\quad (Z=\dot{\phi},\; z:=\dot{\psi}).
\]
 In each of the 3-dimensional subspaces $\Sigma_\Phi$ and $\Sigma_\Psi$ the corresponding projections of the Einstein hypersurface represent hyperbolic ($\Sigma_\Phi$) and elliptic $\Sigma_\Psi$ paraboloids with main axis $O\Phi$ and $O\Psi$,  respectively. All phase trajectories of the dynamical system lie on these \emph{simply connected} hypersurfaces. The inverse transformation to the original potentials, due to its bijectivity, does not violate the simple connectedness of the hypersurface.  This circumstance is the key one that distinguishes our model from the model of \cite{Leon18}, in connection with which the topology of the Einstein - Higgs surface has been given great attention in our work. Note that it would be interesting to apply the methods developed in our article to the quintom model, in which the potential energy is a 6th degree polynomial with respect to the field potentials.  In the future, we intend to conduct a study of models with direct interaction of the components of the quintum, noting that the classification of the topological properties of an algebraic surface of the 6th order, of course, will be immeasurably more difficult to carry out.

Summarizing the brief results of the study, let us highlight its main results.
\begin{enumerate}
\item A rich variety of types of behavior of cosmological models based on the asymmetric Higgs scalar doublet $\Phi, \varphi$, determined by both fundamental parameters and initial conditions, is revealed.
\item A qualitative analysis of the corresponding dynamical system showed that, depending on the values of the fundamental parameters a dynamical system can have 2, 4, 8, 12, 16 or 18 singular points, the character of which is determined by the values of these fundamental parameters of the scalar doublet and the cosmological constant.
\item The concept of an Einstein - Higgs hypersurface of the phase space of a dynamical system is introduced and a classification of its topological properties in three-dimensional subspaces of the classical, $\Sigma_\Phi$, and phantom, $\Sigma_\varphi$, components of the scalar doublet is carried out.
\item A close connection was revealed between the types of behavior of the cosmological model based on an asymmetric scalar doublet and topological
properties of the Einstein - Higgs hypersurface in the three-dimensional subspaces $\Sigma_\Phi$ and $\Sigma_\varphi$.
\item Subclasses of exact stationary solutions of a dynamical system corresponding to singlet fields located in stable
points of the system. These solutions correspond to different cases of expansion ($H=H_0>0$) or contraction ($H=H_0<0$) at a constant rate ($\Omega=1$).
\item The main types of behavior of cosmological models are determined depending on the number and nature of the singular points of the system and topological properties of the Einstein - Higgs hypersurface. Examples of the corresponding types of cosmological models are given and their physical characteristics are studied.
\item It is indicated that the standard models based on an asymmetric scalar doublet correspond to the presence of attractive singular points in regions of positive values of the Hubble constant $H$, repulsive points in the region of negative values of $H$, on the one hand, and large values of the potentials of scalar fields, on the other. In this case, the cosmological model tends monotonically either to an expansion $H>0$ with a constant velocity from below: $\dot {H} \leqslant0$  ($\Omega \to1 $),  or to a contraction $H<0$ with a constant velocity from above ; $\dot{H} \geqslant0 $ ($\Omega \to1 $).
\item In other cases, the cosmological model can make transitions from the expansion mode ($H>0$) to the compression mode ($H<0$), and vice versa. In this case, the transition through the state $H=0$ is accompanied by the tendency of the effective energy of the scalar doublet to zero $\mathcal{E}_m\to 0$ and infinite bursts of cosmological acceleration $\Omega \to \pm \infty$, at which all dynamical variables models remain finite and continuous.
\item Among the models with transitions between expansion and contraction modes, models with transitions between expansion at a constant rate are defined and compression at a constant rate, and vice versa.
\item Among the models with transitions between expansion and contraction modes, models with multiple transitions, including models with an oscillatory mode, have been identified.
\item Among the models with transitions between the expansion and contraction modes, the models with violation and preservation of the sign of the potential of the phantom field, locked in one of the parts of the Einstein - Higgs hypersurface projection breaking in the course of evolution, are defined.
\item The conducted research revealed a special role of the phantom field - ensuring a stable expansion of the cosmological model. In
absence of a phantom field, most cosmological models with small values of the classical scalar field would inevitably slide down
to the compression phase.
\end{enumerate}

Note that the presence of an oscillatory regime in the early stages of the Universe would inevitably lead to the generation of entropy in the early Universe (see, for example, \cite{Zeld}), which could, in principle, fill the missing link in cosmology. Note that in the article \cite{Saridakis}, precisely in connection with the possibility of obtaining such a cosmological scenario that would allow avoiding the singularity of the Big Bang, the prospects of a quintom cosmology were linked. In addition, we note that, contrary to popular belief, the phantom field plays the role of a stabilizer of stable accelerated expansion, thereby being a necessary additional component of the cosmological model.  This property of a seemingly deliberately unstable field to stabilize a dynamical system is quite often manifested in multicomponent essentially nonlinear dynamical systems.

\subsection*{Funding}
The work is performed according to the Russian Government Program  of     Competitive  Growth  of  Kazan  Federal  University.


\begin{thebibliography}{100}
%
\bibitem{Ignat83_1}
Yu.G.~Ignat'ev, ``Conservation laws and thermodynamic equilibrium in the general relativistic kinetic theory of inelastically interacting particles'',  {\it Sov. Phys. J.}, {\bf 26} (1983), 1068.
%
\bibitem{Ignat_Kuz84}
Yu.G.~Ignat'ev, R.R. Kuzeev, ``Thermodynamic equilibrium of a self-gravitating plasma with scalar interaction '', {\it Ukr. Fyz. Journal}, {\bf 29:7} (1984), 1021.
%
\bibitem{Ignat_Mif06}
Yu.G. Ignatyev, R.F. Miftakhov, ``Statistical systems of particles with scalar interaction in cosmology'', {\it Gravitation \& Cosmology}, {\bf 12:2-3} (2006), 179-185.
%
\bibitem{Bron1}
K.A.~Bronnikov, J.C. Fabris, ``Regular Phantom Black Holes'', {\it Phys. Rev. Lett.}, {\bf 96} (2006), 973.
%
\bibitem{Bron2}
S.V.~Bolokhov,  K.A.~Bronnikov, M.V.~Skvortsova, ``Magnetic black universes and wormholes with a phantom scalar'', {\it Clas. and Quant. Grav.}, {\bf 29} (2012), 245006.
%
\bibitem{Ignat_12_3_Iz}
Yu.G. Ignatyev {Ignat'ev}, ``Cosmological evolution of the plasma with interparticle scalar interaction. III. Model with attraction of like-charged scalar particles'', {\it Russian Physics Journal}, {\bf 55:11} (2013), 1345-1350.
\bibitem{Cline}
J.M. Cline, S. Jeon, G.D. Moore, ``The phantom menaced: constraints on low-energy effective ghosts'', {\it Phys.Rev. D70} (2004), 043543; arXiv: hep-ph/0311312.
%
%
\bibitem{Linde}
R. Kallosh, J. Kang, A. Linde, V. Mukhanov, ``The new ekpyrotic ghost'', {\it Journal of Cosmology and Astroparticle Physics}, {\bf 2008} (2008). 10.1088/1475-7516/2008/04/018.
%
\bibitem{Saridakis}
S. Nojiri, E.N. Saridakis, ``Phantom without ghost'', {\it Astrophys Space Sci}, {\bf 347:1}  (2013), 221-226.
%
\bibitem{Sbisa}
F. Sbisa, ``Classical and quantum ghosts'', {\it Europ. J. of Phys.}, {\bf 36:1} (2014), 015009.
%
\bibitem{Vernov}
S.Yu. Vernov, ``Exact solutions of nonlocal nonlinear field equations in cosmology'', {\it Theoretical and Mathematical Physics}, {\bf 166:3} (2011), 392-402.
%
\bibitem{Trodden}
S.M. Carroll, M. Hoffman, M. Trodden, ``Can the dark energy equation-of-state parameter w be less than -1?'', {\it arxiv.org/abs/astro-ph/0301273}, (2003).
%
\bibitem{Richarte}
M. Richarte, G. Kremer, ``Cosmological perturbations in transient phantom inflation scenarios'', {\it The European Physical Journal C}, {\bf 77} (2016), 51.
%
\bibitem{Tripathi}
A. Tripathi, A. Sangwana, H.K. Jassal, ``Dark energy equation of state parameter and its evolution at low redshift'', {\it JCAP}, {\bf 2017:06} (2017), 012.
%
\bibitem{Ma}
Y. Ma, J. Zhang, S. Cao et al., ``The generalized cosmic equation of state: a revised study with cosmological standard rulers'',  {\it Eur. Phys. J. C}, {\bf 77} (2017), 891.
%
\bibitem{Meyers}
J. Meyers et al, ``The Hubble Space Telescope Cluster Supernova Survey. III. Correlated Properties of Type Ia Supernovae and Their Hosts at 0.9 < z < 1.46'', {\it  ApJ}, {\bf 750} (2012).
%
\bibitem{Terlevich}
R. Terlevich,  E. Terlevich  J. Melnick  R. Chavez  M. Plionis  F. Bresolin  S. Basilakos, ``On the road to precision cosmology with high-redshift Hii galaxies'', {\it Monthly Notices of the Royal Astronomical Society}, {\bf 451:13} (2015), 3001-3010.
%
\bibitem{Chavez}
R. Chavez, M. Plionis, S. Basilakos, R. Terlevich, E. Terlevich, J. Melnick, F. Bresolin, A.L. Gonzalez-Moran, ``Constraining the dark energy equation of state with Hii galaxies'', {\it Monthly Notices of the Royal Astronomical Society}, {\bf 462:3} (2016), 2431-2439.
\bibitem{Lazkoz}
R. Lazkoz, G. Leon, ``Quintom cosmologies admitting either tracking or phantom attractors'', Phys.Lett. \textbf{B638} 303 (2006);
arXiv:astro-ph/0602590.
\bibitem{Ignat_12_1_Iz}
Yu.G. Ignat'ev, ``Cosmological evolution of plasma with scalar interparticle interaction. I. Canonical formulation of classical scalar interaction'', {\it Russian Physics Journal},  {\bf 55:2} (2012), 166-172.
\bibitem{Ignat_12_2_Iz}
Yu. G. Ignatiev (Ignat'ev), ``Cosmological evolution of the degenerated plasma with interparticle scalar interaction. II. Formulation of mathematical model'', {\it Russian Physics Journal}, {\bf 55:5} (2012), 550-560.
%
\bibitem{Ignat_14_1_stfi}
Yu.G. Ignat'ev, ``Non-minimal macroscopic scalar field models based on microscopic dynamics '', {\it Space, Time and Fundamental Interactions}, {\bf 1(6)}, (2014), 47-69.
%
\bibitem{Ignat_Dima14_2_GC}
Yu.G.Ignatyev (Ignat'ev), D.Yu.Ignatyev, ``Statistical Systems with Phantom Scalar Interaction in Gravitation Theory. I. Microscopic Dynamics'', {\it Gravitation and Cosmology}, {\bf 20:4} (2014), 299–303.
%
\bibitem{Ignat_Agaf_Dima14_3_GC}
Yu.G. Ignatyev, A.A. Agathonov, and D. Yu. Ignatyev, ``Statistical Systems with Phantom Scalar Interaction in Gravitation Theory. II. Macroscopic Equations and Cosmological Models'', {\it Gravitation and Cosmology}, {\bf 20:4} (2014), 304–308.
%
\bibitem{Ignat_15_1_GC}
Yu.G. Ignatyev (Ignat'ev), ``Nonminimal Macroscopic Models of a Scalar Field Based on Microscopic Dynamics: Extension of the Theory to Negative Masses'', {\it Gravitation and Cosmology}, {\bf 21:4} (2015), 296–308.
%
\bibitem{Ignat_Agaf15_2_GC}
Yu.G. Ignatyev (Ignat'ev), A.A. Agathonov, ``Numerical Models of Cosmological Evolution of a Degenerate Fermi-System of Scalar Charged Particles'', {\it Gravitation and Cosmology}, {\bf 21:2} (2015), 105–112.
%
\bibitem{Ignat_Mih15_2_Iz}
Yu.G. Ignat,ev, M.L. Mikhailov, ``Cosmological Evolution of a Boltzmann Plasma with Interparticle Phantom Scalar Interaction. I. Symmetric Cases'', {\it Russian Physics Journal}, {\bf 57:12} (2015), 1743-1752.
%
\bibitem{Ignat_Agaf_Mih_Dima15_3_AST}
Yu.~Ignat'ev, A.~Agathonov, M.~Mikhailov, D.~Ignatyev, ``Cosmological evolution of statistical system of scalar charged particles'', {\it Astr. Space Sci.}, {\bf 357:61} (2015).
%
\bibitem{Ignat_Agaf_16_3}
Yu.~Ignat'ev, A.~Agathonov, ``Statistical cosmological systems of almost degenerate scalar charged fermions'', {\it Space, Time and Fundamental Interactions}, {\bf 3} (2016), 48-90.
%
\bibitem{Ignat_Sasha_G&G}
Yu.G. Ignat'ev, A.A. Agathonov, and D.Yu. Ignatyev, ``Statistical Cosmological Fermion Systems with Phantom Scalar Interaction of Particles,'' {\it Gravitation and Cosmology}, {\bf 24} (2018), 1-12.
%
\bibitem{8}
Yu.G. Ignat'ev,  ``Non-equilibrium kinetic models of the Universe. I. Conditions of local thermodynamic equilibrium'', {\it Space, Time and Fundamental Interactions}, {\bf 1} (2012), 79-98.
 %
 \bibitem{12}
u.G. Ignat,ev, M.L. Mikhailov, ``Cosmological Evolution of a Boltzmann Plasma with Interparticle Phantom Scalar Interaction. I. Phase transitions: a simplified model '', {\it Space, Time and Fundamental Interactions}, {\bf 4} (2015), 75-90.
%
\bibitem{Ignat_16_5_Iz}
Yu.G. Ignatyev, ``Qualitative and Numerical Analysis of the Cosmological Model with a Phantom Scalar Field'', {\it Russian Physics Journal}, {\bf 59:12} (2017), 2074–2079.
%
\bibitem{Ignat_Agaf_16_6_stfi}
Yu.~G.~Ignat'ev and  A.~A.~Agathonov, ``Qualitative and Numerical Analysis of a Cosmological Model Based on a Phantom Scalar Field with Self-Interaction'', {\it Space, Time and Fundamental Interactions}, {\bf 4} (2016), 52-61.
%
\bibitem{Ignat_Agaf_2017_2_GC}
Yu.~G.~Ignat'ev and  A.~A.~Agathonov, ``Qualitative and Numerical Analysis of a Cosmological Model Based on a Phantom Scalar Field with Self-Interaction'', {\it Grav. and Cosmol.}, {\bf 23:3} (2017), 230-235.
%
\bibitem{Bogoyav}
 O. I. Bogoyavlensky, ``The methods of the qualitative theory of dynamic systems in astrophysics and gas dynamics'', Moskow, Nauka (1980).
%
\bibitem{Bautin}
N.N. Bautin, E.A. Leontocich, ``Methods and techniques for the qualitative study of dynamical systems on a plane '', Moskow, Nauka (1989).
%
%
\bibitem{Yu_STFI_3}
Yu.~G.~Ignat'ev, ``Qualitative and numerical analysis of cosmological models based on an asymmetric scalar doublet: classical + phantom scalar field. I. The case of minimally interacting scalar fields: a qualitative analysis'', \emph{Space, Time and Fundamental Interactions}, \textbf{2} (2017), 36-52.
 %
%
\bibitem{DifEqTools}
Yu.G. Ignat'ev, A.R. Samigullina,  ``Numerically - analytical methods of mathematical modeling of nonlinear dynamic systems in SCM Maple. III. Visualization of nonlinear dynamical systems and applications to mechanics and cosmology'', \emph{Space, Time and Fundamental Interactions}, \textbf{3} (2017), 28-54.
%
\bibitem{Ignat_Sam2}
Yu.G. Ignat'ev, A.R. Samigullina, ``Averaging of the Equations of the Standard Cosmological Model over Rapid Oscillations: Influence of the Cosmological Term on the on the mean value of the effective barotropic coefficient'', \emph{Russian Physics Journal}, \textbf{61:4} (2018), 643-647.
%
\bibitem{Vernov1}
I.Ya. Aref’eva, A.S. Koshelev, and S.Yu. Vernov, ``Crossing the $w=?1$ barrier in the D3-brane dark energy model'', \emph{Phys. Rev. D}, \textbf{72:6} (2005), 064017; arXiv: astro-ph/0507067.
%
\bibitem{Vernov2}
I.Ya. Aref’eva, S.Yu. Vernov, A.S. Koshelev, ``Exact solution in a string cosmological model'', \emph{Theor. Math. Phys.}, \textbf{148:1} (2006), 895–909; arXiv:astro-ph/0412619.	
%
\bibitem{Vernov3}
S.Yu. Vernov, ``Construction of exact solutions in two-field cosmological models'', \emph{Theor. Math. Phys.}, \textbf{155:1} (2008),  544–556  (2008)
%
\bibitem{fantom}
R.R. Caldwell, ``A phantom menace? Cosmological consequences of a dark energy component with super-negative equation of state'', \emph{Phys. Lett. B}, \textbf{545:1–2} (2002), 23-29.
%
\bibitem{Yu_17_1}
Yu.G. Ignat’ev, ``Qualitative and Numerical Analysis of a Cosmological Modely Based on a Classical Massive Scalar Field'', \emph{Grav. Cosmol}., \textbf{23:2} (2017), 131–141.
%
\bibitem{Yu_19_1}
Yu.G. Ignat’ev, I.A. Kokh,	``Peculiarities of Cosmological Models Based on a Nonlinear Asymmetric Scalar Doublet with Minimal Interaction. I. Qualitative Analysis'',	\emph{Gravit. and Cosmol}., \textbf{25:1} (2019),  24–36.
%
\bibitem{Yu_19_2}
Yu.G. Ignat’ev, I.A. Kokh,	``Peculiarities of Cosmological Models Based on a Nonlinear Asymmetric Scalar Doublet with Minimal Interaction. II. Numerical Analysis'',	\emph{Grav. Cosmol}., \textbf{25:1} (2019),  37–43.
%
\bibitem{Yu_19_3}
Yu.G. Ignat'ev, A.R. Samigullina,	``On euclidean limit cycles in cosmological models based on scalar fields'', \emph{Russian Physics Journal}, \textbf{62:4}  (2019), 618-626.
%
\bibitem{Yu_20_1}
Yu.G. Ignat’ev, D.Yu. Ignatyev,	``A Complete Model of Cosmological Evolution of a Scalar Field with Higgs Potential and Euclidean Cycles'',	\emph{Grav. Cosmol}., \textbf{26:1} (2020), 29–37.
%
\bibitem{Leon18}
Genly Leon (Catolica del Norte U.), Andronikos Paliathanasis, Jorge Luis Morales, ``The past and future dynamics of quintom dark energy models'', Eur. Phys. J. C 78, 753 (2018).
%
\bibitem{Zeld}
Ja.B. Zeldovich, I.D. Novikov, \emph{The structure and evolution of the universe}, Moskow, Nauka, 1975.
%
\bibitem{Saridakis}
Yi-Fu Cai, Emmanuel N. Saridakis, Mohammad R. Setare, Jun-Qing Xia, Quintom Cosmology: Theoretical implications and observations.   Phys.Rept.493:1-60, 2010; 	arXiv:0909.2776 [hep-th].
%
\end{thebibliography}
\end{document}